\documentclass[epj]{svjour}
\usepackage{epsfig,array}
\usepackage{multirow}
\usepackage{graphicx}
\newcommand{\be}{\begin{eqnarray}}
\newcommand{\ee}{\end{eqnarray}}

\newcommand{\nn}{\nonumber }
\newcommand{\bea}{\begin{eqnarray}}
\newcommand{\eea}{\end{eqnarray}}

\newcommand{\beq}{\begin{equation}}
\newcommand{\eeq}{\end{equation}}

\newcommand{\KL}{K^+\rm\Lambda}
\newcommand{\KS}{K^+\rm\Sigma^0}

\def\fun#1#2{\lower3.6pt\vbox{\baselineskip0pt\lineskip.9pt
\ialign{$\mathsurround=0pt#1\hfil##\hfil$\crcr#2\crcr\sim\crcr}}}

\begin{document}

\title{Baryon resonances and polarization transfer in hyperon
photoproduction}

\author{
A.V.~Anisovich \inst{1,2}
\and V.~Kleber \inst{3}
\and E.~Klempt \inst{1}
\and V.A.~Nikonov   \inst{1,2}
\and A.V.~Sarantsev \inst{1,2}
\and U.~Thoma \inst{1}
}

\institute{
Helmholtz--Institut f\"ur Strahlen-- und Kernphysik,
Universit\"at Bonn, Germany
\and Petersburg Nuclear Physics Institute, Gatchina, 188300 Russia
\and Physikalisches Institut, Universit\"at Bonn, Germany
}

\date{Received: \today / }

\abstract{A partial wave analysis of data on
photoproduction of hyperons including single and double polarization
observables is presented. The large spin transfer probability reported by the CLAS
collaboration can be successfully described within an
isobar partial wave analysis.}
\authorrunning{A.V.~Anisovich {\it et al.}}
\titlerunning{Transfer of polarization in hyperon photoproduction}
\mail{klempt@hiskp.uni-bonn.de}

\maketitle

\section{Introduction}

The new CLAS data on hyperon photoproduction \cite{Bradford:2006ba}
show a remarkably large spin transfer probability. In the reactions
$\gamma p\to \Lambda K^+$ and $\gamma p\to\Sigma K^+$ using a
circularly polarized photon beam, the polarizations of the $\Lambda$
and $\Sigma$ hyperons were monitored by measurements of their decay
angular distributions. For photons with helicity $h_{\gamma}=1$, the
magnitude of the $\Lambda$ polarization vector was found to be close
to unity, $1.01\pm0.02$ when integrated over all production angles
and all center-of-mass energies $W$. For $\Sigma$ photoproduction,
the polarization was determined to be $0.82\pm0.03$ (again
integrated over all energies and angles), still a remarkably large
value. The polarization was determined from the expression
$\sqrt{C_x^2 + C_z^2 + P^2}$, where $C_z$ is the projection of the
hyperon spin onto the photon beam axis, $P$ the spin projection on
the normal-to-the-reaction plane, and $C_x$ the spin projection in
the center-of-mass frame onto the third axis. The measurement of
polarization effects for both $\Lambda$ and $\Sigma$ hyperons is
particularly useful. The $ud$ pair in the $\Lambda$ is antisymmetric
in both spin and flavour; the $ud$ quark carries no spin, and the
$\Lambda$ polarization vector is given by the direction of the spin
of the strange quark. In the $\Sigma$ hyperon, the $ud$ quark is in
a spin-1 state and points into the direction of the $\Sigma$ spin
while the spin of the strange quark is opposite to it.

Schumacher \cite{Schumacher:2006} interpreted the $\gamma p\to
\Lambda K^+$ process on the quark level. He assumed that the
circularly polarized photon with $h_{\gamma}=1$ converts into a
$\phi$-meson due to vector-meson dominance; the $s\bar s$ pair would
have helicity 1 and the spins of the $s$-quarks would then be
transferred to the hyperon spin. Since the $s\bar s$ pair is created
in the initial state, $s$-channel $N^*$ baryons cannot play a large
role in the reaction, and the data should be explained with
dominantly non-resonant contributions to the reaction amplitude.
However, it is well known that resonances do play a significant
role, especially in the $\KL$ channel. Furthermore, if the
strange-quark helicity were responsible for the $\Lambda$
polarization, we should expect opposite polarizations for $\Lambda$
and $\Sigma$. This is however forbidden by selection rules: for
forward and backward angles, $C_x$ is constrained to zero and $C_z$
to unity ($+1$). The model \cite{Schumacher:2006} fails to reproduce
the qualitative features of $\gamma p\to\Sigma K^+$.

Kochelev  \cite{Kochelev:2006fn} argues that instantons provide a
natural explanation of the strong polarization transfer. Instanton
induced interactions transform initial-state quarks with left-handed
(right-handed) chiralities into final-state quarks having opposite
chiralities. For massless quarks, chirality coincides with helicity,
so the helicities of the initial and final state should be fully
correlated. It is again difficult to understand why the $\Lambda$
and the $\Sigma$ spin are both aligned with the photon polarization

The recoil (or induced) polarization $P$ can be orien\-ted
pa\-rallel or anti-parallel to the scattering-plane normal. Thus,
the recoil polarizations $P_{\Lambda}$ and $P_{\Sigma}$ can have
opposite signs. An opposite sign for $\Lambda$ and $\Sigma$
polarizations has been observed in a number of reactions; induced
polarization is well established and not restricted to
photoproduction reactions \cite{Wilkinson:1981jy}. When the
$\Lambda$ and $\Sigma$ polarizations in a given process point into
the opposite directions and have the same magnitude, this may serve
as evidence that the hadronic polarization originates from the
underlying orientation of the strange quark spin. DeGrand and
Miettinen \cite{DeGrand:1981pe} have argued that in the presence of
a strong gradient of the interaction potential, there is an
effective $L\cdot S$ interaction which increases (or decreases) the
scattering angle for $L\cdot S>0$ (or $L\cdot S<0$) and, thus, an
effective
\begin{figure*}[pt]
\vspace{3mm} \centerline{
\epsfig{file=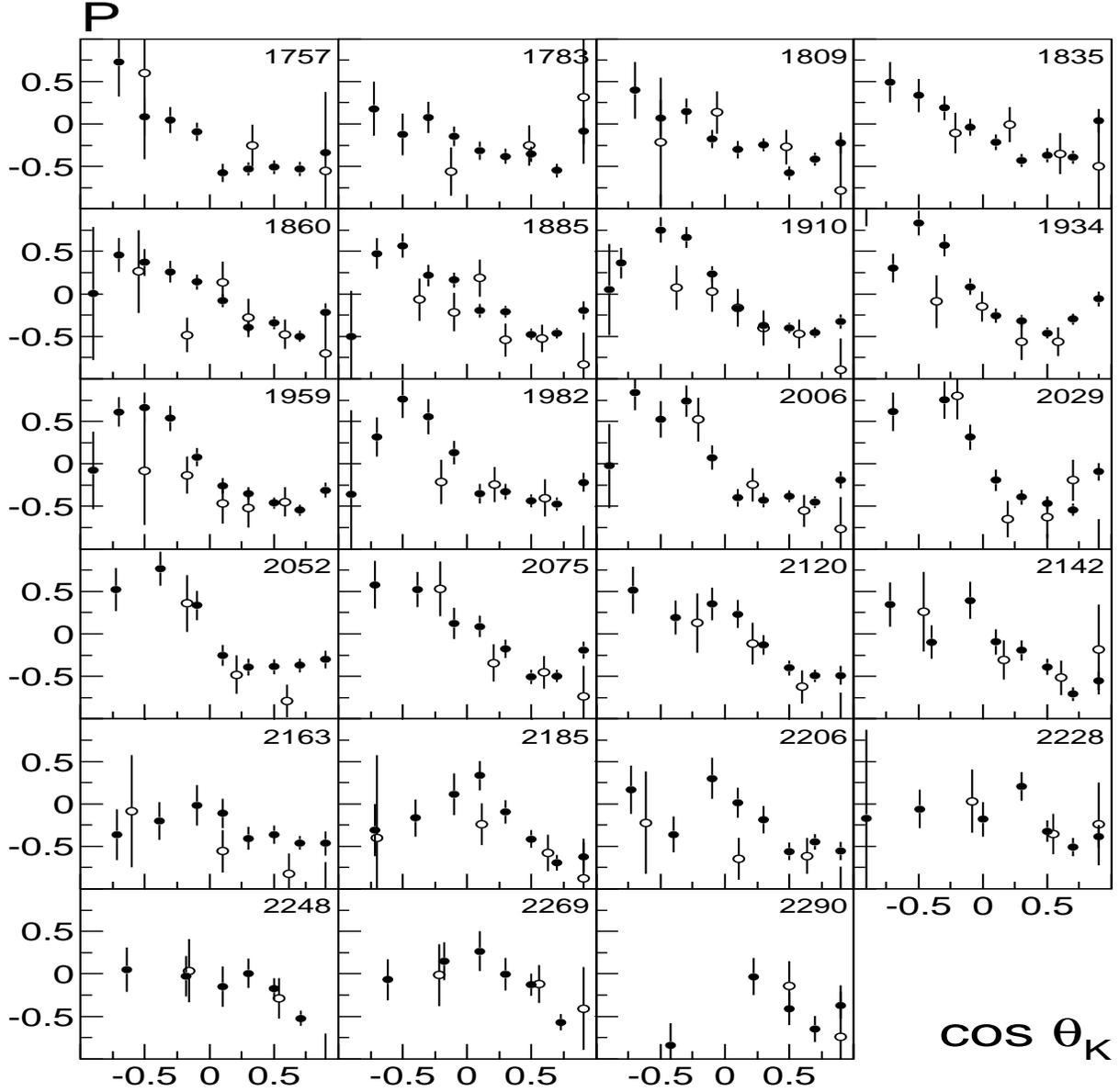,width=0.90\textwidth,height=0.68\textheight}}
\caption{\label{P-Lam-Sig} The recoil polarizations $P_{\Lambda}$
for $\gamma p \rightarrow \KL$ (black circles), and $P_{\Sigma}$ for
$\gamma p \rightarrow \KS$ (open circles) \cite{Bradford:2005pt}.
The latter is plotted with a negative sign. }
\end{figure*}
polarization asymmetry is induced. The $L\cdot S$
interaction can be experienced by the $s$ quark at the moment of
$s\bar s$ pair creation, or at the hadron level.

Fig. \ref{P-Lam-Sig} shows the CLAS data \cite{Bradford:2005pt} on the
induced polarization $P_{\Lambda}$ and, overlaid, $-P_{\Sigma}$. The
induced polarizations $P_{\Lambda}$ and $P_{\Sigma}$ evidently have
opposite signs and have similar magnitudes.

Independently of the question if the polarization phenomena require
an interpretation on the quark or on the hadron level, the large
polarization seems to contradict an isobar picture of the process in
which intermediate $N^*$'s and $\Delta^*$'s play a dominant role. It
is therefore important to see if the data are compatible with such
an isobar interpretation or not.

\section{Double polarization observables in hyperon photoproduction}

Let us briefly summarize the definition and properties of the
beam-recoil observables $C_x$ and $C_z$. The coordinate system
$(x,y,z)$ is defined by the normal to the reaction plane $y$ and
the direction of the incoming photon $z$. Following the notation given
in \cite{Fasano:1992es}, the polarization cross section can be written
as
\be
\sigma^{(B, T, R)} \equiv \frac{d\sigma^{(B, T, R)}}{d\Omega}
\ee
where the subscript $(B, T, R)$ denotes the polarization sta\-tes of
the beam, target and recoil baryon, respectively. To express the
beam-recoil observables $C_i$, we use a rotated coordinate system with
$y'=y$ and $z'$ defined by the direction of the outgoing kaon. The
photon beam with helicity $h_{\gamma}=+1$ is marked by the subscript
$r$. The beam-recoil observables $C_{x'}$ and $C_{z'}$
are then defined as
\be C_{x'} = \frac{\sigma^{(r, 0, +x')} -
\sigma^{(r, 0, -x')}} {\sigma^{(r, 0, +x')} + \sigma^{(r, 0, -x')}}
\nonumber \\ C_{z'} = \frac{\sigma^{(r, 0, +z')} - \sigma^{(r, 0,
-z')}} {\sigma^{(r, 0, +z')} + \sigma^{(r, 0, -z')}}
\ee
The properties of $C_{x'}$ and $C_{z'}$ are well known, its expressions
in terms of CGLN (Chew-Goldberger-Low-Nambu) \cite{Chew:1957tf},
helicity and multipole amplitudes can be found for example in
\cite{Fasano:1992es}. The CLAS collaboration measured this beam-recoil
observables in the system where the z axis of the reaction plane is
along the direction of the photon beam. Since the polarization is
transformed like a three-vector, the transition from $C_{x'}$,$C_{z'}$
to $C_x$,$C_z$ follows the standard rotation matrix, \be
C_x&=&C_{x'}\cos \theta +C_{z'}\sin\theta \nn \\
C_z&=-&C_{x'}\sin \theta +C_{z'}\cos\theta
 \label{cx_cz}
\ee
where $\theta$ is the scattering angle of the kaon. The
$C_{x'}$,$C_{z'}$ or $C_x$,$C_z$ observables have defined values at
forward and backward scattering angles. Helicity conservation
requires $C_x = 0$ and $C_z = 1$ at $\theta = 0,180$. This feature
is model independent and, of course, compatible with the CGLN
amplitudes. Hence, $C_z = 1$ holds for both, for the photoproduction
of $\KL$ and of $\KS$.

The double polarization observables $C_x$ and $C_z$ have an important
constraint given by \cite{Barker:1975bp}
\be C_{x}^2 + C_{z}^2 \leq\;
Min((1 -\Sigma^2), (1-P^2)) \label{limits}
\ee
where $\Sigma$ is the beam asymmetry and $P$ the recoil polarization.
Evidently, there is a strong connection between $C_i$ and single
polarization observables.

\section{The partial wave formalism}

The partial wave analysis presented here is based on relati\-vistically
invariant amplitudes constructed from the four-momenta of particles
involved in the process \cite{Anisovich:2004zz}. High-spin resonances
are described by relativistic multi-channel Breit-Wigner amplitudes,
important partial waves with low total spin ($J<5/2$) are described in
the framework of the K-matrix/P-vector approach which automatically
satisfies the unitarity condition. The amplitude for elastic scattering
is given by
 \be
\hat A(s) \;=\; \hat K\;(\hat I\;-\;i\hat \rho \hat K)^{-1}\,.
\ee

The phase space matrix $\hat \rho$ is a diagonal matrix. The two
particle phase space was used in the form calculated in
\cite{Anisovich:2006bc}:
\be
\label{psr}
\rho(s)=\frac{\alpha_L}{2L+1}
\frac{2|\vec k|^{2L+1}}{\sqrt{s}}\frac{k_{10}+m_N}{2m_N}
\frac {F(\vec k^2)}{B(L,r,\vec k^2)}
\ee
where $s$ is the total energy squared, $k$ is the relative momentum
between baryon and meson, $k_{10}$ is the energy of the baryon (with
mass $m_N$) calculated in the c.m.s. of the reaction and the
coefficient $\alpha_L$ is equal to:
\be
\alpha_L=\prod\limits_{n=1}^L\frac{2n-1}{n}\,.
\ee
For regularization of the phase volume at large energies we used a
standard Blatt-Weisskopf form factor with $r=0.8$ fm and a form factor
$F(k^2)$ using two different forms:
\be
F(k^2)=\frac{\Lambda+0.5}{\Lambda+\vec k^2}\,,\qquad
F(k^2)=\frac{\Lambda+2.5}{\Lambda+s}\,,
\ee
Fits with both parameterizations  produced a very similar result.
The parameter $\Lambda$ was fixed from our previous analysis
\cite{Anisovich:2005tf,Sarantsev:2005tg}  and was
not varied in the present fit. For the first parameterization it was
taken to be equal to 1.5 and for the second one 3.0. Below threshold
the two body phase volumes were continued with the subtracted
dispersion integral:
\be
\label{psi} i\rho(s)\;\to
 \int\limits_{(m_N\!+\!m_\mu)^2}^{\infty}\!\!\!\!
\frac{\rho(s')(s-(m_N\!+\!m_\mu)^2)}{(s'-s)(s'-(m_N\!+\!m_\mu)^2)}\frac{ds'}\pi\
,
\ee
 where $m_\mu$ is the meson mass in the decay channel. The exact
formulas for the three body phase volumes are given in
\cite{Anisovich:2006bc}.

The photoproduction amplitude can be written in the P-vector
approach. The P-vector amplitude is then given by
\be
 A_a \;=\; \hat P_b\;(\hat I\;-\;i\hat \rho \hat
K)^{-1}_{ba}\,.
\ee
 The production  vector $\hat P$ and the K-matrix
$\hat K$ have the following parameterizations:
 \be
 K_{ab}\;=\;\sum_\alpha \frac{g_a^{(\alpha)} g_b^{(\alpha)}}
{M^2_\alpha - s} \;+\; f_{ab},
 \;\;\;
 P_{b}\;=\;\sum_\alpha \frac{
g_{\gamma \rm N}^{(\alpha)} g_b^{(\alpha)}}{M^2_\alpha - s} \;+\;
 \tilde f_{b} \nonumber\\
\label{Pvect}
 \ee
 where $M_\alpha$, $g_a^{(\alpha)}$ and $g_{\gamma\rm N}^{(\alpha)}$
 are the mass, coupling constant and photo-coupling of the resonance
 $\alpha$; $f_{ab}$ describes direct non-resonant transition processes
 from an initial state $a$ to a final state $b$, e.g. from $N\pi\to\Lambda K$ The production process
 may have a non-resonant contribution due to  $\tilde f_{b}$. In general,
 these non-resonant contributions are functions of $s$.

 For all partial waves except $S_{11}$, it is sufficient to assume
$f_{ab}$ and $\tilde f_{b}$ to be constants. Only the $S_{11}$ wave
requires a slightly more complicated structure. For the scattering
amplitudes $\pi N\to N\pi$, $\pi N\to N\eta$, and $\eta N\to N\eta$
we choose
\be
f_{ab} =\frac{f_{ab}^{(1)}+f_{ab}^{(2)}\sqrt s}{s-s_0^{ab}}\,.
\ee
This form is similar to the one used by SAID \cite{Arndt:2006bf}.
The $f_{ab}^{(i)}$ and $s_0^{ab}$ are constants which are determined
in the fits.

The P-vector approach is based on the idea that a channel with a
weak coupling can be omitted from the K-matrix. Indeed, adding to the
K-matrix the $\gamma N$ channel would not change  the properties of
the amplitude. Due to its weak coupling, the $\gamma N$ interaction can
be taken into account only once; this is done in the form of a
P-vector. Loops due to virtual decays of a resonance into $N\gamma$ and
back into the resonance can be neglected safely. A similar approach can
be used to describe also decay modes with a weak coupling. The
amplitude for the transition into such a channel can be written as
D-vector amplitude, \be
\label{amplitude}
 A_a \;=\; \hat D_a + [\hat K
(\hat I\;-\;i\hat \rho \hat K)^{-1}\,\hat \rho ]_{ab} \hat D_{b}\;,
\ee
\begin{table}[pt]
\caption{\label{Table:p11_km} Properties of the $P_{11}$ $K$-matrix
poles for one of the solutions. The masses, widths, and $g$ are
given in GeV, the helicities $A$ in GeV$^{-\frac{1}{2}}$, $s$ in
GeV$^2$; the $f$ are dimensionless.}
\begin{center}
\renewcommand{\arraystretch}{1.4}
\begin{tabular}{clcccc}
\hline\hline
&                      &Pole 1       &Pole 2       &Pole 3       &        \\
\hline
& $M$                  &  1516$\pm$30&  1722$\pm$35&  2233$\pm$50&       \\
& $A_{1/2}$            & -0.031      &  0.022      & -0.067      &       \\
\hline
   $ a $&              &$g^{(1)}_a$  &$g^{(2)}_a$  &$g^{(3)}_a$  &$f_{1a}$\\
\hline
 1&$N(940)\pi $         &  1.052      & -0.156      &  1.226      & -1.736\\
 2&$\Lambda K^+$        &  0          & -0.003      & -0.556      &  0    \\
 3&$\Sigma  K^+$        &  0          &  0.709      &  0.063      &  0    \\
\hline\hline
\end{tabular}
\renewcommand{\arraystretch}{1.0}
\end{center}
\caption{\label{Table:p13_km_1} Properties of the $P_{13}$
$K$-matrix poles for one of the solutions. }
\begin{center}
\renewcommand{\arraystretch}{1.4}
\begin{tabular}{clcccc}
\hline\hline
 &                     &Pole 1       &Pole 2       &Pole 3       &        \\
\hline
 &$M$                  &  1770$\pm$50&  1880$\pm$50&  2195$\pm$50&
 \\ &$A_{1/2}$            &  0.022      &  0.010      &  0.202      &
 \\ &$A_{3/2}$            &  0.051      &  0.039      & -0.069      &
 \\
\hline
   $ a $&              &$g^{(1)}_a$  &$g^{(2)}_a$  &$g^{(3)}_a$  &$f_{1a}$\\
\hline
 1&$N(940)\pi $         &  0.650      &  0.206      & -1.200      & -1.237   \\
 2&$N(940)\eta $        &  0.900      &  0.811      & -0.440      & -0.962   \\
 3& $\Lambda K^+$        &  0.900      &  0.560      & -0.849      & -0.016   \\
 4&$\Sigma  K^+$        & -0.319      &  0.727      & -0.202      &  1.050   \\
\hline\hline
\end{tabular}
\renewcommand{\arraystretch}{1.0}
\end{center}
\caption{\label{Table:s11_km_1} Properties of the $S_{11}$
$K$-matrix poles for one of the solutions. } \begin{center}
\renewcommand{\arraystretch}{1.4}
\begin{tabular}{clcccc}
\hline\hline
&               &Pole 1       &Pole 2       &          &  \\
\hline & $M$           &  1420$\pm$80&  1710$\pm$30&          & \\
&$A_{1/2}$  &  0.125      &  0.093      &          & \\
\hline
   $ a $ &      &$g^{(1)}_a$  &$g^{(2)}_a$  &$f_{1a}^{(1)}$&$f_{2a}^{(1)}$ \\
\hline
 1&$N(940)\pi $  &  0.074      &  0.844      &  1.651      & -1.970   \\
 2&$N(940)\eta $ & -1.908      & -0.434      & -1.970      &  2.010   \\
 3&$\Lambda K^+$ & -0.300      & -0.346      & -0.016      & -0.432   \\
 4&$\Sigma  K^+$ & -1.232      & -0.804      & -0.497      &  3.663   \\
\hline
 &              &$f_{1\alpha}^{(2)}$&$s_0^{1a}$&$f_{2a}^{(2)}$&$s_0^{2a}$ \\
\hline
 1&$N(940)\pi $  & -1.285      &  0.430      &  0.625      & -0.620   \\
 2&$N(940)\eta $ &  0.625      & -0.620      &  1.990      &  0.600   \\
\hline\hline
\end{tabular}
\renewcommand{\arraystretch}{1.0}
\end{center}
\end{table}
where the parameterization of the D-vector is similar to the
parameterization of the P-vector in (\ref{Pvect}). As in the case of
the P-vector approach, channels with weak couplings can be taken
into account only in their real decay, not as virtual loops.

In case where both initial and final coupling constants are weak, we
used the so-called PD-vector approximation. In this case the
amplitude is given by
\be
 A_{ab} \;=\; \hat G_{ab} + \hat P_{a}
(\hat I\;-\;i\hat \rho \hat K)^{-1}\,\hat \rho \hat D_{b}\;,
\ee
where $\hat G$ corresponds to direct production from state $'a'$
decaying into state $'b'$. The P-vector and D-vectors are defined
above.

To facilitate the interpretation of the formalism we give in Tables
\ref{Table:p11_km}--\ref{Table:s11_km_1} the full parametrization of
the K-matrices for the $P_{11}$, $P_{13}$, and $S_{11}$ waves for
one of the two solutions discussed below.

\section{Data and fits}
The data used in this analysis comprise differential cross section
for $\gamma p\to \KL$, $\gamma p\to \KS$, and $\gamma p\to
\Sigma^+K^0_s$ including their recoil polarization  and the photon
beam asymmetry, and the recent spin transfer measurements
\cite{Bradford:2006ba,Bradford:2005pt,Glander:2003jw,McNabb:2003nf,%
Lleres:2007tx,Zegers:2003ux,Lawall:2005np,Castelijns:2007qt}.
Furthermore, data are included from the SAID data base \cite{Bartholomy:04,%
Bartalini:2005wx,SAID1,SAID,SAID2,Crede:04,Krusche:nv,GRAAL2,%
Bartalini:2007fg} on photoproduction of $\pi^0$ and $\eta$ with
measurements of differential cross sections, beam and target
asymmetries and recoil polarization. We did not use the $\KS$ recoil
polarization data from \cite{Lleres:2007tx} since they have larger
errors and a smaller energy range than the CLAS data.

The fit also used data on photo-induced $2\pi^0$ production
\cite{Thoma:2007,Sarantsev:2007} and $\pi^0\eta$ \cite{Horn:2007}
and the recent BNL data on $\pi^- p\to n\pi^0\pi^0$
\cite{Prakhov:2004zv} in an event-based likelihood fit.
$2\cdot\ln\mathcal L$ was added to the pseudo-$\chi^2$ function. The
data essentially determined the contributions of isobars to the
$N\pi\pi$ and $N\pi\eta$ final state and are not discussed here
further. Details can be found in
\cite{Thoma:2007,Sarantsev:2007,Horn:2007}.

\begin{table}[pt] \caption{\label{list_chi} Single meson
photoproduction data used in the partial wave analysis and $\chi^2$
for solutions 1 and 2.}
\renewcommand{\arraystretch}{1.2}
\begin{center}
\begin{tabular}{lccccr}
\hline\hline
&&&&&\vspace*{-3mm}\\
Observable\hspace*{-3mm}&\hspace*{-3mm}$N_{\rm data}$\hspace*{-3mm}&
\hspace*{-3mm}$w_i$\hspace*{-3mm}&\hspace*{-3mm}$\frac{\chi^2}{N_{\rm data}}$
\hspace*{-4mm}&\hspace*{-4mm}$\frac{\chi^2}{N_{\rm data}}\hspace*{-3mm}$&
 Ref.\vspace*{2mm} \\
\hline \hline
& &&Sol. 1 &  Sol. 2& \\
$C_x(\rm\gamma p \rightarrow \Lambda K^+)$\hspace*{-3mm}
 &   160 &5&    1.71  &  1.66   &
\cite{Bradford:2006ba} \\
$C_z(\rm\gamma p \rightarrow \Lambda K^+)$\hspace*{-3mm}
 &   160 &7&    1.95  &  2.34    &
\cite{Bradford:2006ba} \\
$\rm\sigma(\gamma p \rightarrow \Lambda K^+)$\hspace*{-3mm}
 &  1377 &5&  2.02  &  1.99     & \cite{Bradford:2005pt} \\
 $\rm\sigma(\gamma p \rightarrow \Lambda K^+)$\hspace*{-3mm}
 &  720 &1&   1.53  &  1.55     &
\cite{Glander:2003jw} \\
$\rm P(\gamma p \rightarrow \Lambda K^+)$\hspace*{-3mm}
 &  202 &6.5&   1.65  &  2.28     &
\cite{McNabb:2003nf} \\
$\rm P(\gamma p \rightarrow \Lambda K^+)$\hspace*{-3mm}
 &  66 &3&   2.89  &  1.05     &
\cite{Lleres:2007tx} \\
$\Sigma(\rm\gamma p \rightarrow \Lambda K^+)$\hspace*{-3mm}
 &   66 &5&    2.19  &  2.85   &
\cite{Lleres:2007tx} \\
$\Sigma(\rm\gamma p \rightarrow \Lambda K^+)$\hspace*{-3mm}
 &   45 &10&    1.98  &  1.82    &
\cite{Zegers:2003ux} \\
\hline
$C_x(\rm\gamma p \rightarrow \Sigma^0 K^+)$\hspace*{-3mm}
 &   94 &5&    2.70  &  3.50     &
\cite{Bradford:2006ba} \\
$C_z(\rm\gamma p \rightarrow \Sigma^0 K^+)$\hspace*{-3mm}
 &   94 &5&    2.77  &  2.24    &
\cite{Bradford:2006ba} \\
$\rm\sigma(\gamma p \rightarrow \Sigma^0 K^+)$\hspace*{-3mm}
 &  1280 &3&  2.10  &  2.19     &
\cite{Bradford:2005pt} \\
$\rm\sigma(\gamma p \rightarrow \Sigma^0 K^+)$\hspace*{-3mm}
 &  660 &1&   1.33  &  1.41     &
\cite{Glander:2003jw} \\
$\rm P(\gamma p \rightarrow \Sigma^0 K^+)$\hspace*{-3mm}
 &   95 &6&   1.58  &  1.94     &
\cite{McNabb:2003nf} \\
$\Sigma(\rm\gamma p \rightarrow \Sigma^0 K^+)$\hspace*{-3mm}
 &   42 &5&    1.04  &  1.34     &
\cite{Lleres:2007tx} \\
$\Sigma(\rm\gamma p \rightarrow \Sigma^0 K^+)$\hspace*{-3mm}
 &   45 &10&    0.62  &  0.76     &  \cite{Zegers:2003ux} \\
\hline
$\rm\sigma(\gamma p \rightarrow \Sigma^+ K^0)$\hspace*{-3mm}
 &   48 &2.3&   3.51  &  3.41    &
\cite{McNabb:2003nf} \\
$\rm\sigma(\gamma p \rightarrow \Sigma^+ K^0)$\hspace*{-3mm}
 &  120 &5&   0.98  &  1.09    &
\cite{Lawall:2005np} \\
$\rm\sigma(\gamma p \rightarrow \Sigma^+ K^0)$\hspace*{-3mm}
 & 72  &5& 1.17    &  0.77    &
\cite{Castelijns:2007qt} \\
\hline
$\rm\sigma(\gamma p \rightarrow p\pi^0)$\hspace*{-3mm}
 & 1106 &7&  0.99  &  1.03    &
\cite{Bartholomy:04} \\
$\rm\sigma(\gamma p \rightarrow p\pi^0)$\hspace*{-3mm}
 &  861 &3&  3.22  &  2.44   &
 \cite{Bartalini:2005wx} \\
$\Sigma(\rm\gamma p \rightarrow p\pi^0)$\hspace*{-3mm}
 &  469 &2.3&  3.75  &  3.35    &
 \cite{Bartalini:2005wx}\\
$\Sigma(\rm\gamma p \rightarrow p\pi^0)$\hspace*{-3mm}
 &  593 &2.3&  2.13  &  2.20    &
 \cite{SAID1}\\
$\rm P(\rm\gamma p \rightarrow p\pi^0)$\hspace*{-3mm}
 &  594 &3&  2.58  &  2.54    &
 \cite{SAID}\\
$\rm T(\rm\gamma p \rightarrow p\pi^0)$\hspace*{-3mm}
 &  380 &3&  3.85  &  3.90    &
 \cite{SAID}\\
$\rm\sigma(\gamma p \rightarrow n\pi^+)$\hspace*{-3mm}
 & 1583 &2.8&  1.07  &  1.27    &
\cite{SAID2} \\
\hline
$\rm\sigma(\gamma p \rightarrow p\eta)$\hspace*{-3mm}
 &  667 &30&   0.84  &  0.77   &
\cite{Crede:04} \\
$\rm\sigma(\gamma p \rightarrow p\eta)$\hspace*{-3mm}
 &  100 &7&   1.69  &  1.97   &
 \cite{Krusche:nv}\\
$\Sigma(\rm\gamma p \rightarrow p\eta)$\hspace*{-3mm}
 &   51 &10&   1.82  &  1.91    &
 \cite{GRAAL2}\\
$\Sigma(\rm\gamma p \rightarrow p\eta)$\hspace*{-3mm}
 &  100 &10&   2.11  &  2.24   &
 \cite{Bartalini:2007fg}\\
\hline $\gamma \to p2\pi^0$ & 160k & 3
&\multicolumn{2}{c}{likelihood fit}&\cite{Thoma:2007}\\
$\gamma \to p\pi^0\eta$ & 16k & 5
&\multicolumn{2}{c}{likelihood fit}&\cite{Horn:2007}\\
$\pi^-p \to n2\pi^0$ & 180k & 2.5-4
&\multicolumn{2}{c}{likelihood fit}&\cite{Prakhov:2004zv}\\
\hline $P_{11}(\pi N\to N\pi)$\hspace*{-3mm}
 & 110  &20 & 1.60    & 1.74    &
\cite{Arndt:2006bf} \\
$P_{13}(\pi N\to N\pi)$\hspace*{-3mm}
 & 134  &10 &  3.78   & 2.83    &
\cite{Arndt:2006bf} \\
$S_{11}(\pi N\to N\pi)$\hspace*{-3mm}
 & 126  &30 & 1.86    & 1.84    &
\cite{Arndt:2006bf} \\
$D_{33}(\pi N\to N\pi)$\hspace*{-3mm}
 & 108  &12 & 1.88    &  2.69   &
\cite{Arndt:2006bf} \\
 \hline \hline
\end{tabular}
\end{center}
\renewcommand{\arraystretch}{1.0}
\vspace{-0.3cm}
\end{table}

Data on $\pi N$ elastic data from the SAID data base
\cite{Arndt:2006bf} were used for those partial waves which are
described by a K-matrix.

Fits were performed using a pseudo-$\chi^2$ function
\be
\chi^2_{\rm tot}=\chi^2_{\rm 2b}\;-\;2\ln\mathcal L\ ,\quad
\chi^2_{\rm 2b}=\frac{\sum w_i\chi^2_i}{\sum w_i\,N_i}\,\sum N_i\ ,
\label{chi_tot}
\ee
where the $N_i$ are given as $N_{\rm data}$ (per channel) in the
second and the weights in the third column of Table~\ref{list_chi}.
The data were fitted with weights $w_i$ which ensure that
low-statistics data are described reasonably well. Without weights,
high-statistics data enforce a very good description, without taking
into account any model imperfections, while low-statistics data
would be reproduced badly in fits.

Table \ref{old} summarizes the resonances used in our fits. In
addition, amplitudes for some further resonances are included which
make minor contributions to photoproduction, $N(1700)D_{13}$,
$N(1710)P_{11}$, $N(1875)D_{13}$, $N(2000)F_{15}$, $N(2170)D_{13}$,
$N(2200)P_{13}$, ~$\Delta(1600)P_{33}$, ~$\Delta(1905)F_{35}$,\\
$\Delta(1920)P_{33}$, $\Delta(1950)F_{37}$, and amplitudes
representing $t$- and $u$-channel exchanges. The $S_{11}$-wave was
fitted as 2-pole 5-channel K-matrix ($\pi N$, $\eta N$, $K\Lambda$,
$K\Sigma$, $\Delta(1232)\pi$); the $P_{11}$-wave as 3-pole 4-channel
K-matrix ($\pi N$, $\Delta(1232)\pi$, $K\Sigma$ and $N\sigma$) and
$D_{33}$-wave as 2-pole 3-channel K-matrix ($\pi N$,
$\Delta(1232)\pi$ ($S$ and $D$-waves)). The $P_{13}$ K-matrix was
approximated as 3-pole 8-channel K-matrix with $\pi N$, $\eta N$,
$\Delta(1232)\pi$ (P- and F-waves), $N\sigma$, $D_{13}(1520)\pi$,
$K\Lambda$ and $K\Sigma$ channels.

\begin{table}[pt]
\caption{\label{old}Baryon resonances used in the fits. All
resonances listed contribute more than 4\% of the intensity to at
least one of the photo-production reactions listed in Table
\ref{list_chi}. The underlined resonances provide a significant
fraction to hyperon photoproduction. Further resonances are needed
to achieve a good fit.}
\renewcommand{\arraystretch}{1.6}
\begin{center}
\begin{tabular}{ccccc}
\hline\hline
\hspace{-1mm}$N(1440)P_{11}$\hspace{-1mm}&\hspace{-1mm}$N(1520)D_{13}$
\hspace{-1mm}&\hspace{-1mm}\underline{$N(1535)S_{11}$}\hspace{-1mm}&
\hspace{-1mm}\underline{$N(1650)S_{11}$}\hspace{-1mm}\\
\hspace{-1mm}$N(1675)D_{15}$\hspace{-1mm}&\hspace{-1mm}$N(1680)F_{15}$
\hspace{-1mm}&
\hspace{-1mm}\underline{$N(1720)P_{13}$}\hspace{-1mm}&\hspace{-1mm}\underline{$N(1840)P_{11}$}
\hspace{-1mm}\\
\hspace{-1mm}\underline{$N(1900)P_{13}$}
&\hspace{-1mm}\hspace{-1mm}$N(2070)D_{15}$ \hspace{-1mm}&&
\hspace{-1mm}$\Delta(1232)P_{33}$\hspace{-1mm}\\
\hspace{-1mm}$\Delta(1620)S_{31}$\hspace{-1mm}&
\hspace{-1mm}$\Delta(1700)D_{33}$\hspace{-1mm} &
\hspace{-1mm}$\Delta(1940)D_{33}$\hspace{-1mm}&$\Delta(1950)F_{37}$\hspace{-1mm}&\\[+0.5ex]
\hline \hline
\end{tabular}
\end{center}
\renewcommand{\arraystretch}{1.0}
\end{table}

 A $\chi^2_{\rm 2b}$ of better than 20.000 was reached for 16.000
data points. In spite of the use of differential cross sections, of
single and double polarization observables, the results of the fits
depended on the starting values. Two separate classes of solutions were
found, giving rather different isobar contributions. These will be
compared in the discussion of the data. The two classes of solutions
will be called solution 1 and 2, respectively.

One new resonance is added to describe the new CLAS data, the
$N(1900)P_{13}$, for which so far, evidence had been weak only. It
is surprising that the new very significant data on $C_x$ and $C_z$
are well described by introducing just one single resonance to the
model. Compared to our previous analysis
\cite{Anisovich:2005tf,Sarantsev:2005tg}, the $\Delta(1600)P_{33}$
and $N(1710)P_{11}$ have been introduced when the data on two-pion
production and the elastic $\pi N$ scattering amplitude were
included in the fit \cite{Thoma:2007,Sarantsev:2007}. Here, just
$N(1900)P_{13}$ was added. We also tried to introduce additional
resonances, one by one, in the $1/2^{\pm}$, 3$/2^{\pm}$,
$5/2^{\pm}$, $7/2^{\pm}$, $9/2^{\pm}$ partial waves, without finding
a significant improvement.

\begin{figure*}[pt] \centerline{
\epsfig{file=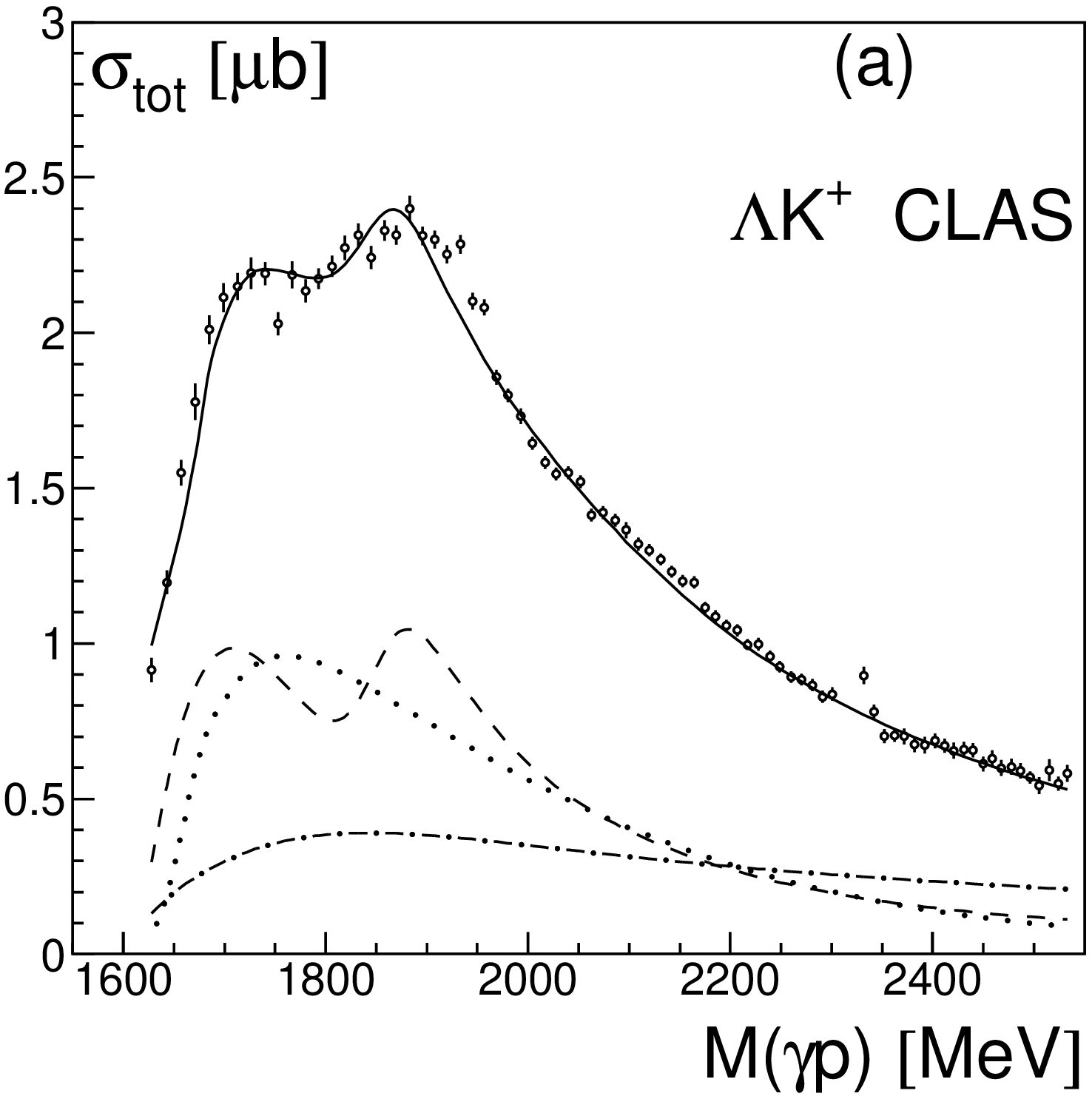,width=0.40\textwidth}
\hspace{4mm}\epsfig{file=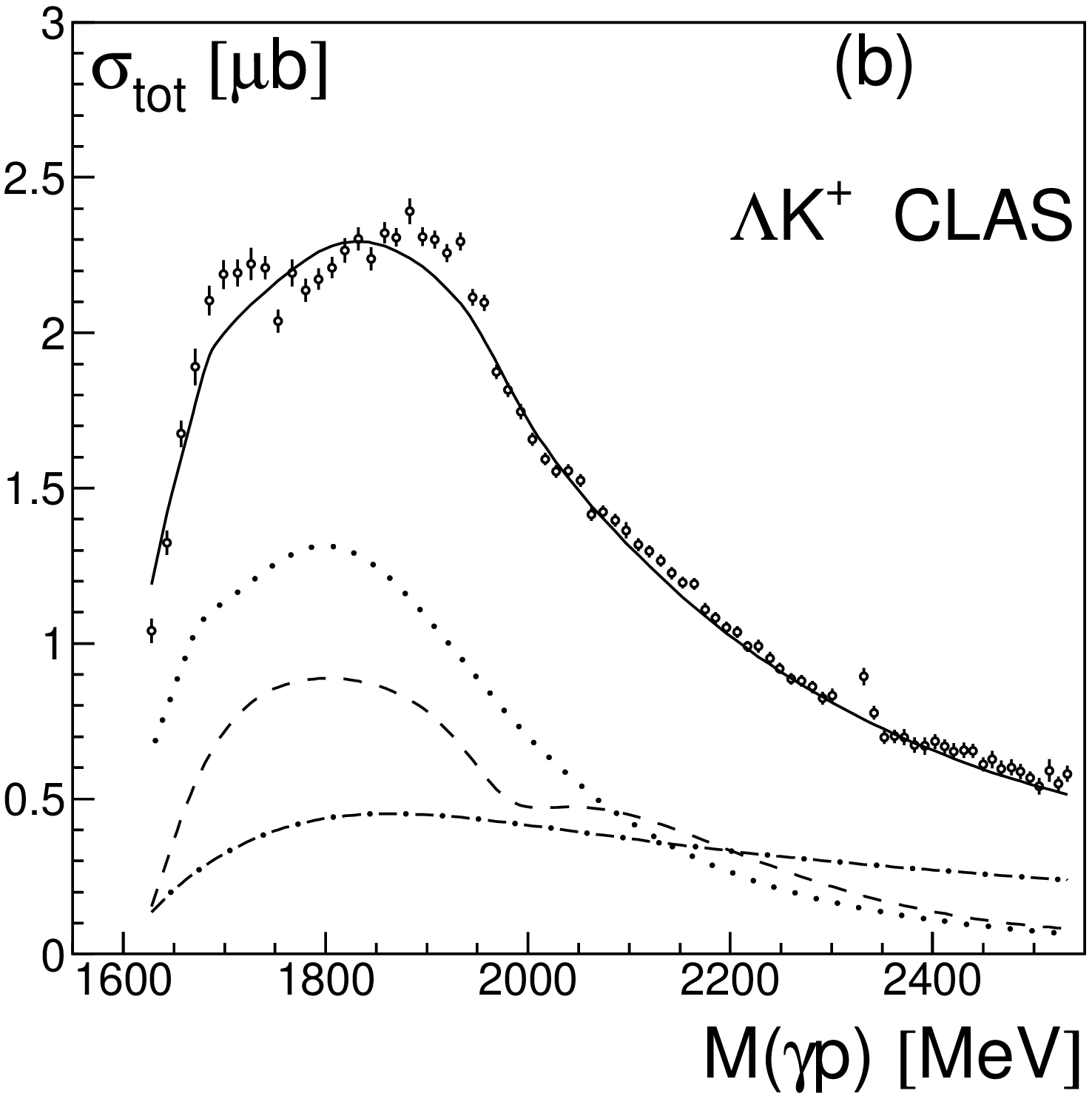,width=0.40\textwidth}
} \caption{\label{tot_klam} The total cross section for $\gamma p
\rightarrow \KL$ \cite{Bradford:2005pt} for solution 1 (a) and
solution 2 (b). The solid curves are the results of our fits, dashed
lines are the $P_{13}$ contribution, dotted lines are the $S_{11}$
contribution and dash-dotted lines are the contribution from $K^*$
exchange. \vspace{3mm}
 }
\centerline{
\epsfig{file=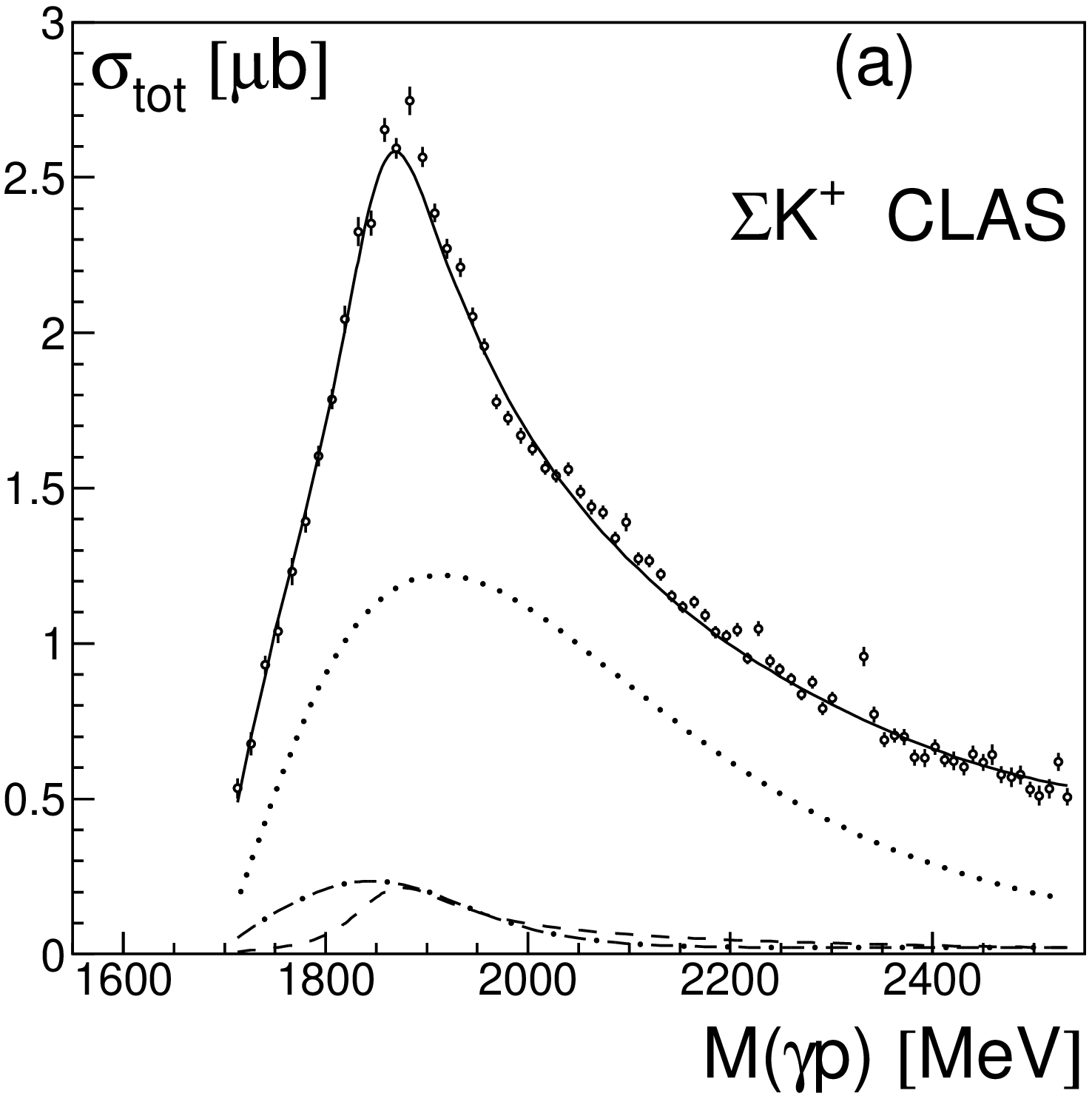,width=0.40\textwidth}
\hspace{4mm}\epsfig{file=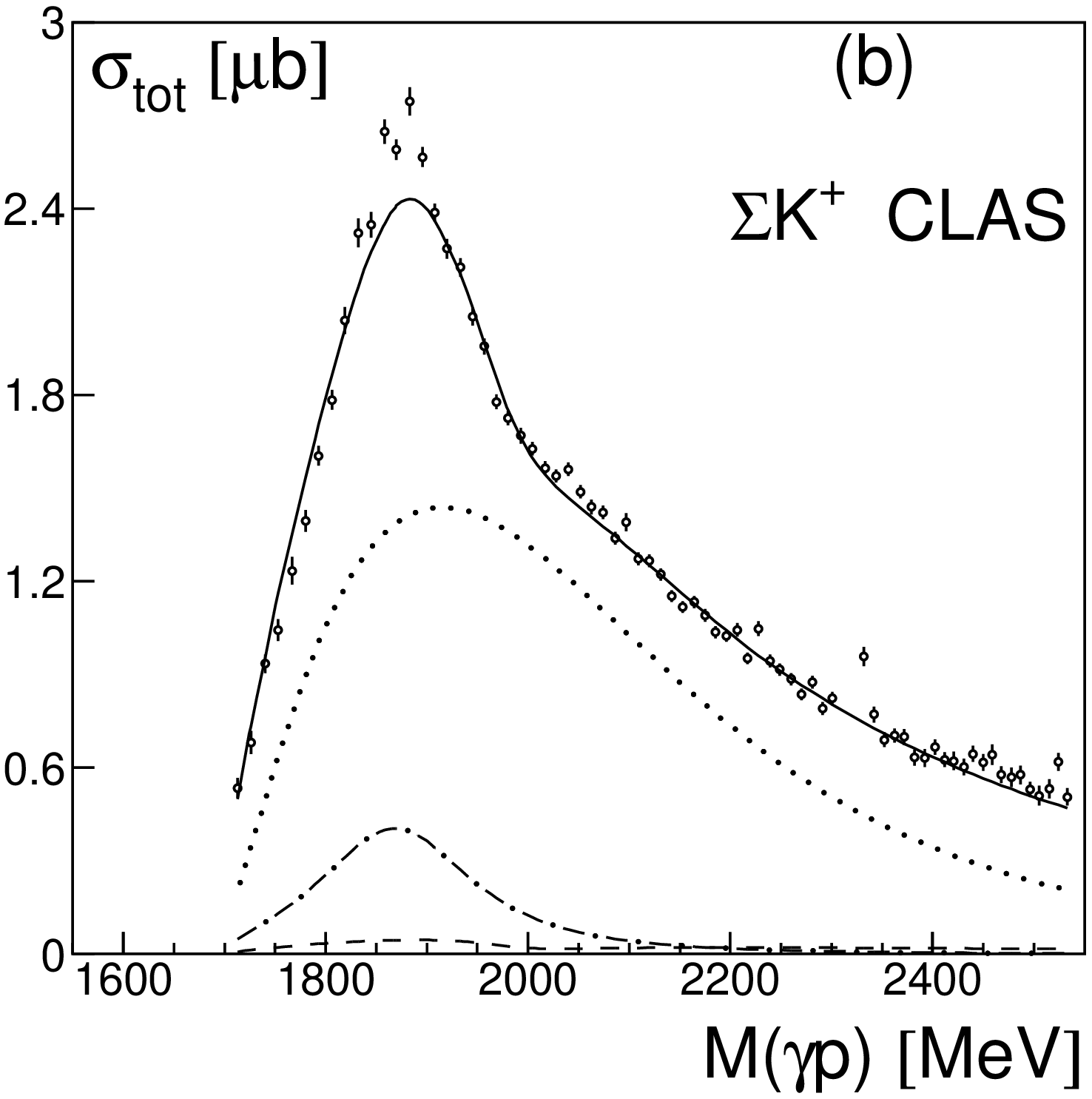,width=0.40\textwidth}
} \caption{\label{ksig_tot} The total cross section for $\gamma p
\rightarrow \KS$ \cite{Bradford:2005pt} for solution 1 (a) and
solution 2 (b). The solid curves are the results of our fits, dashed
lines are the $P_{13}$ contribution, dash-dotted lines are the
$P_{31}$ contribution and dotted lines are the contribution from $K$
exchange.
 }
\end{figure*}
\begin{figure*}[pt]
\begin{tabular}{cc}
\hspace{-2mm}\epsfig{file=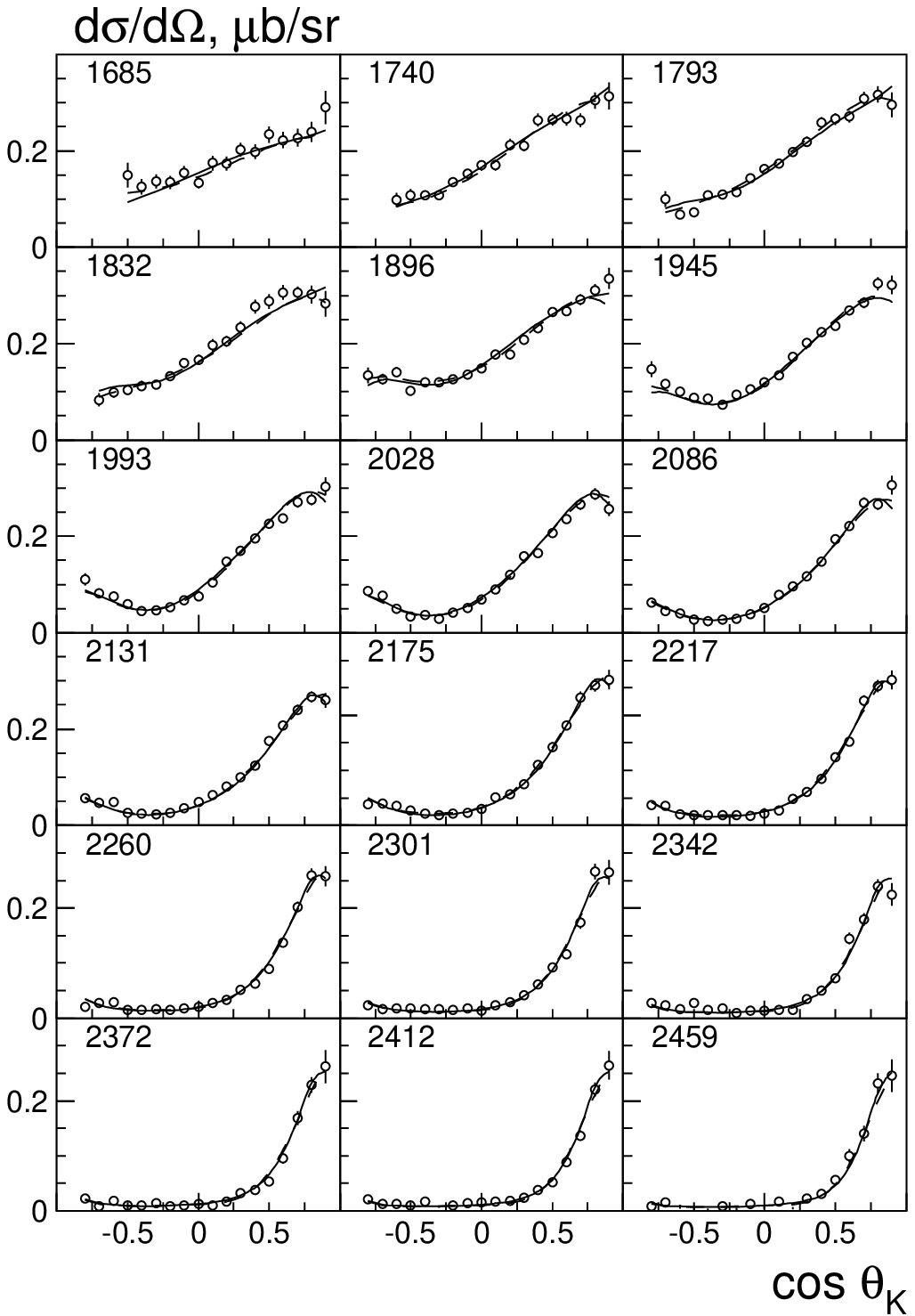,width=0.50\textwidth,height=0.48\textheight}&
\hspace{-4mm}\epsfig{file=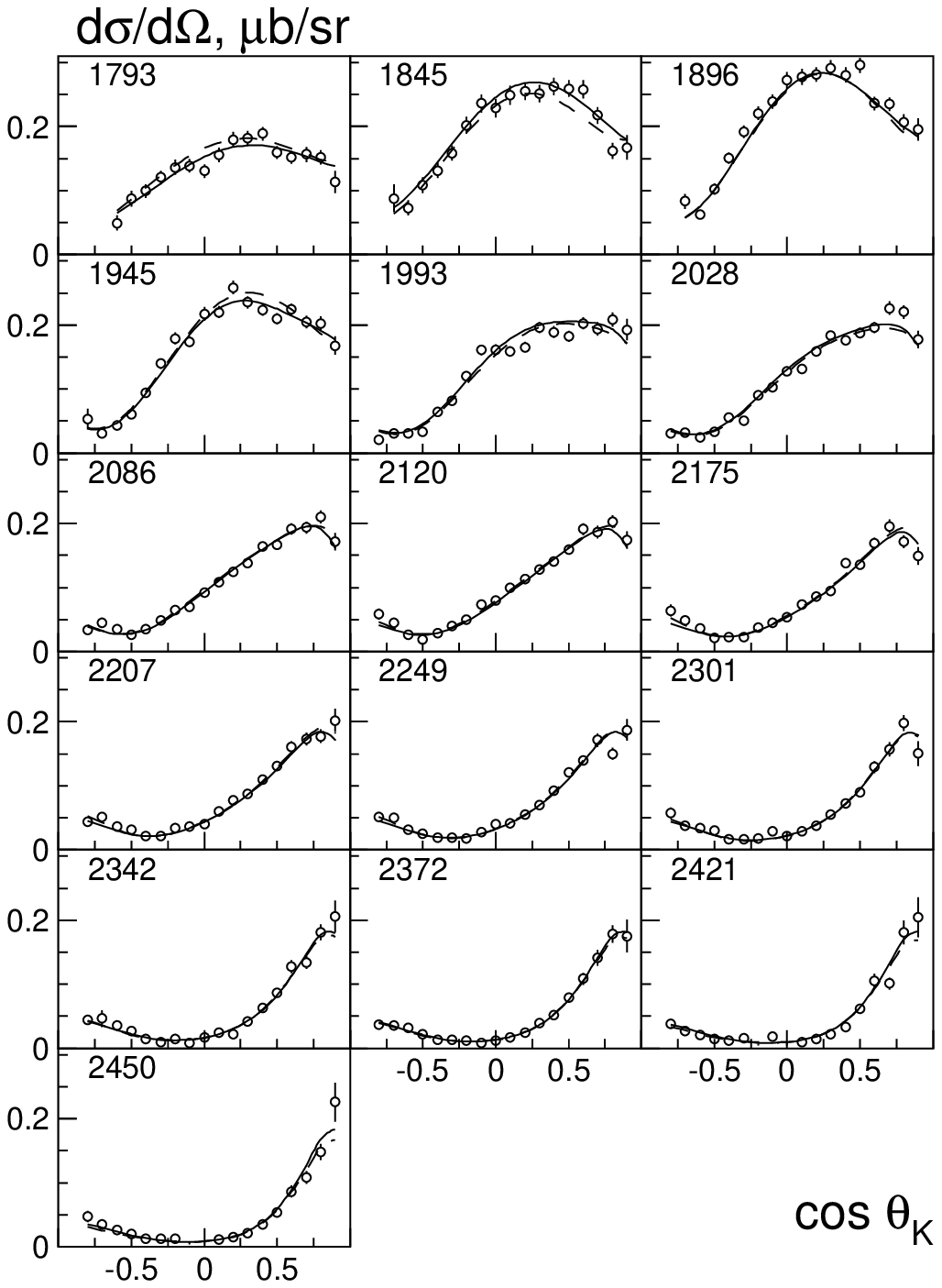,width=0.50\textwidth,height=0.48\textheight}
\end{tabular}
\caption{\label{dcs_klam}Differential cross sections for $\gamma p
\rightarrow \KL$ (left) and $\gamma p \to \KS$ (right)
\cite{Bradford:2005pt}. Only energy bins where $C_x$ and $C_z$ were
measured are shown. The solution 1 is shown as solid line and solution
2 (hardly visible since overlapping) as a dashed line. The total energy
is given in MeV.}
\begin{tabular}{cc}
\hspace{-2mm}\epsfig{file=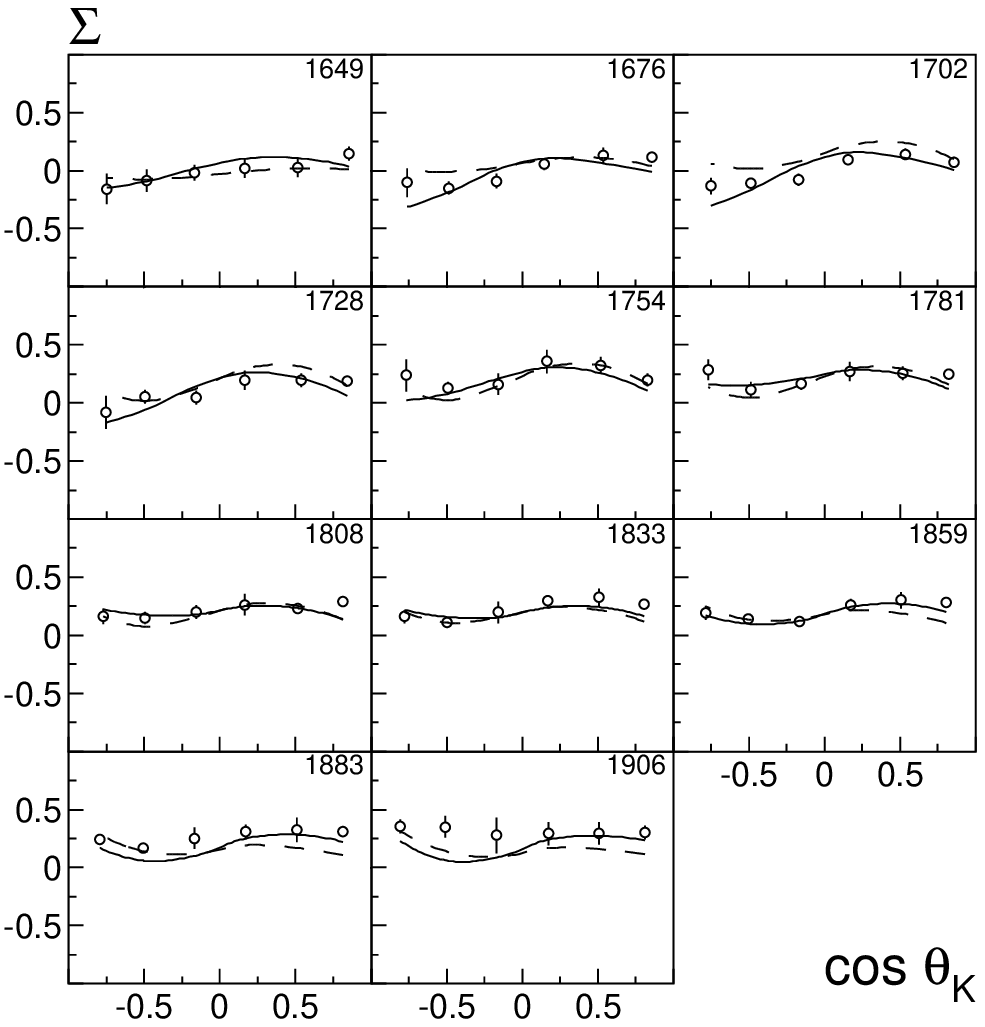,width=0.50\textwidth,height=0.42\textheight}&
\hspace{-4mm}\epsfig{file=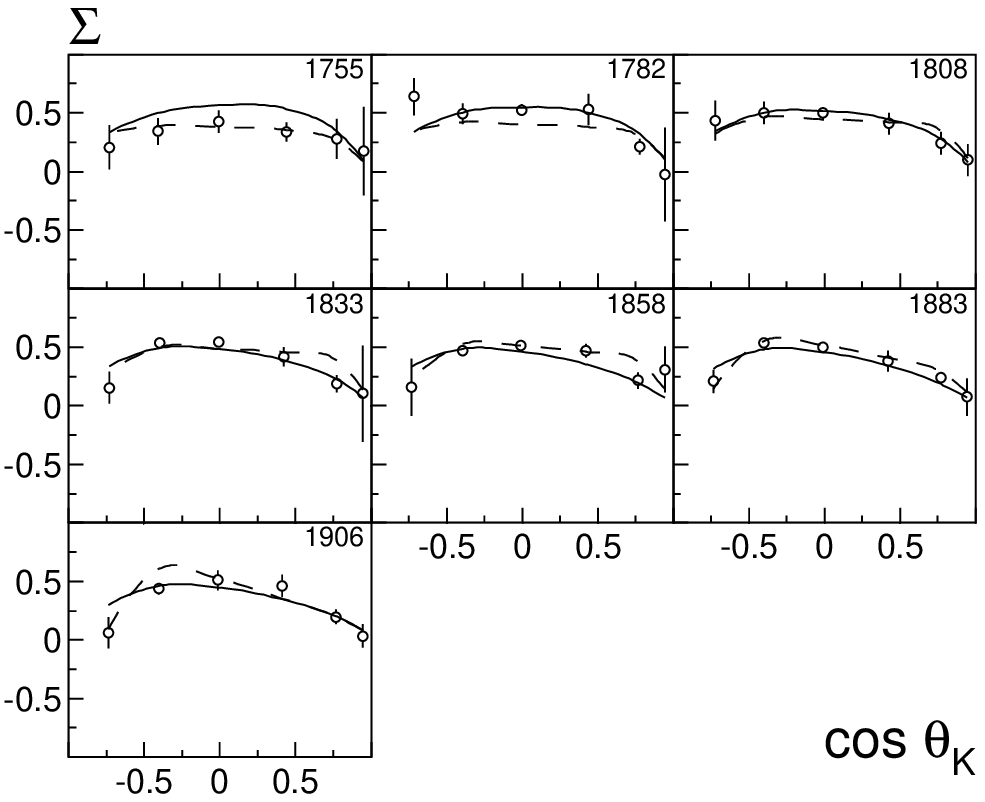,width=0.50\textwidth,height=0.42\textheight}
\end{tabular}
\caption{\label{fig:klam_graal06_s} The beam asymmetries as a
function of $W$ for $\rm \gamma p\to K^{+}\Lambda$ (left) and $\rm
\gamma p\to K^{+}\Sigma$ (right) \cite{Lleres:2007tx}. The solid and
dashed curves are the result of our fit obtained with solution 1 and
2, respectively.}
\end{figure*}

\begin{figure*}[pt]
\begin{center}
\begin{tabular}{cc}
\hspace{-2mm}\epsfig{file=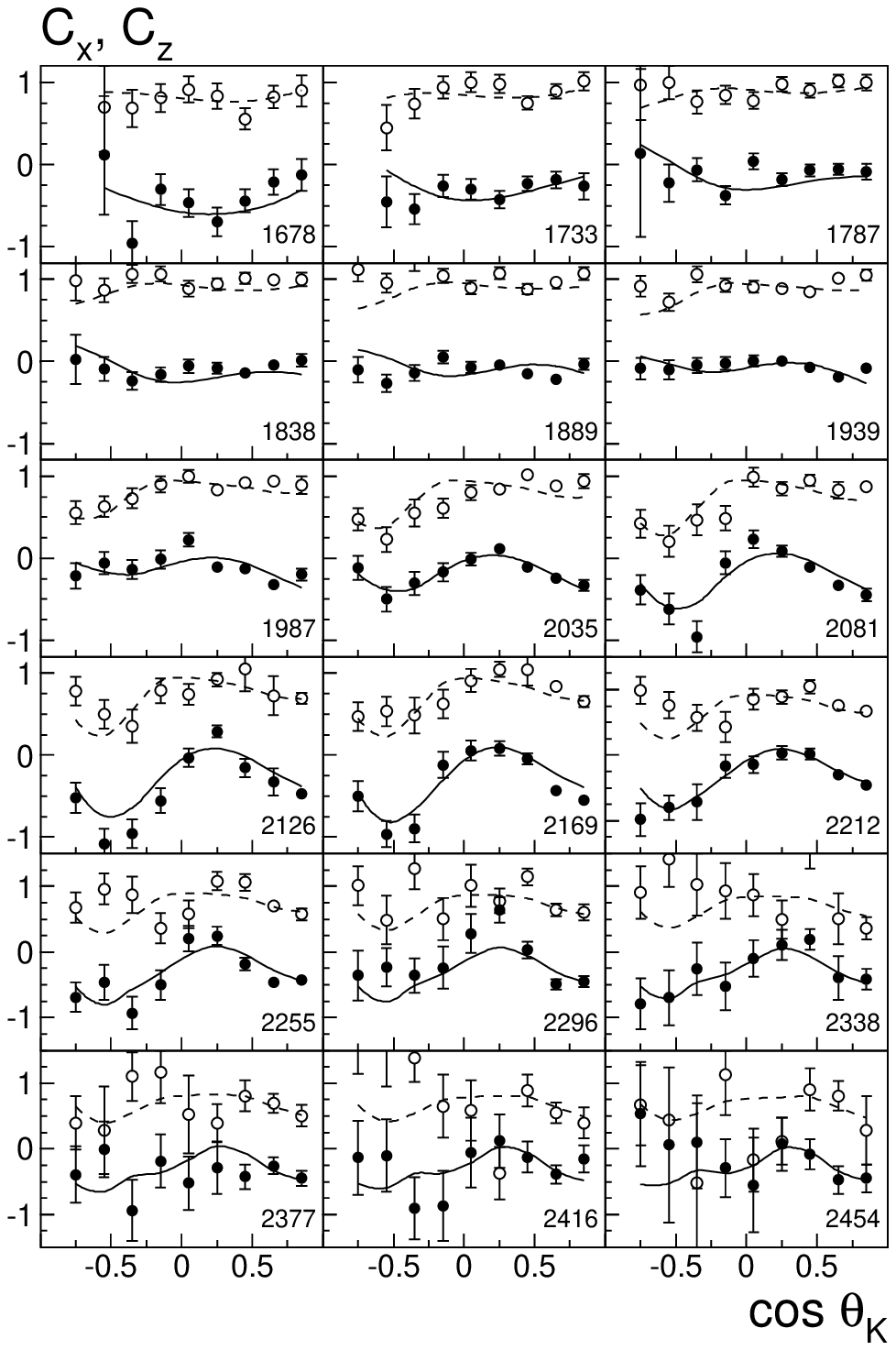,width=0.50\textwidth,height=0.47\textheight}&
\hspace{-4mm}\epsfig{file=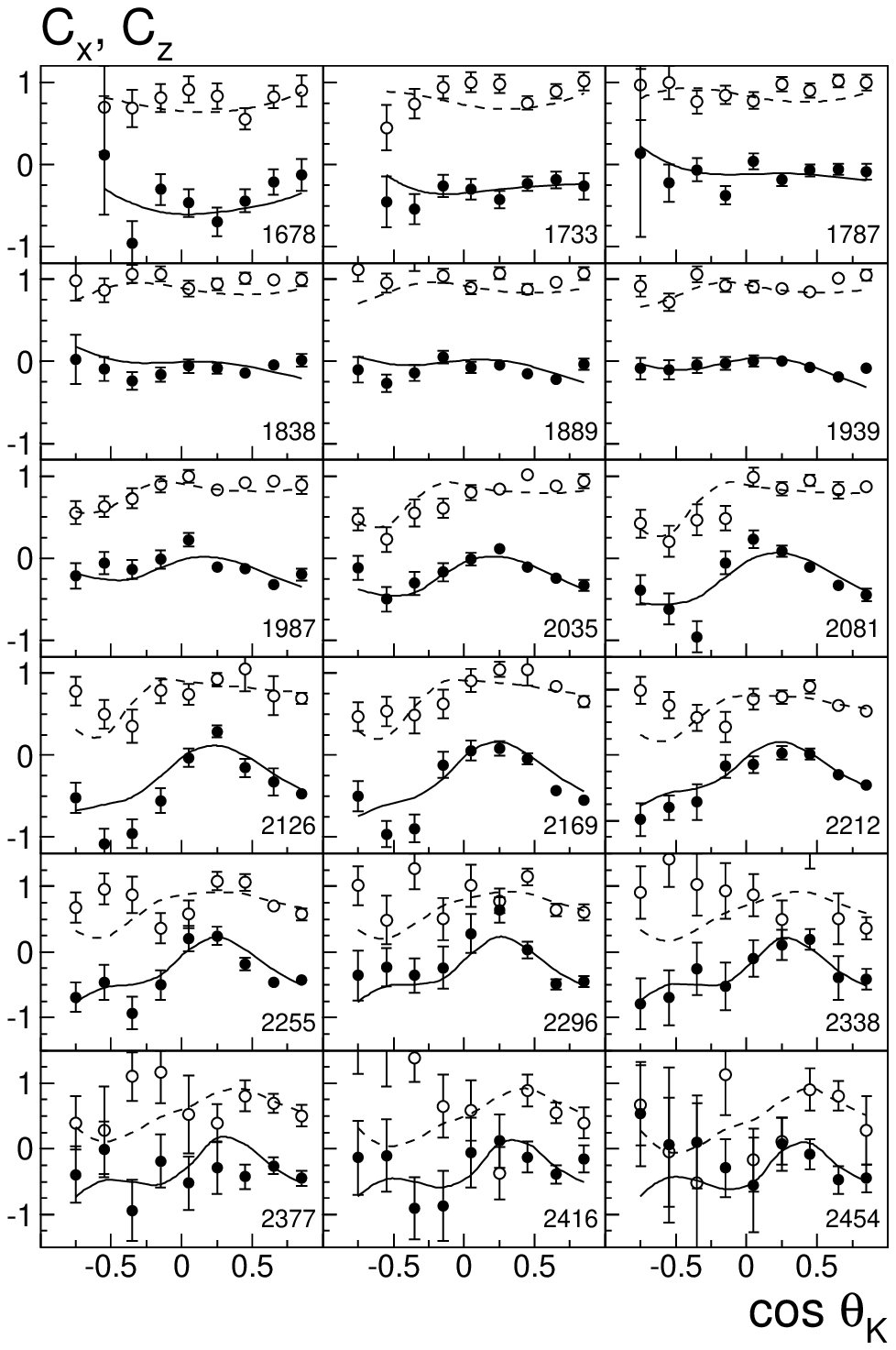,width=0.50\textwidth,height=0.47\textheight}
\end{tabular}
\vspace{-4mm}
\begin{tabular}{cc}
\hspace{-2mm}\epsfig{file=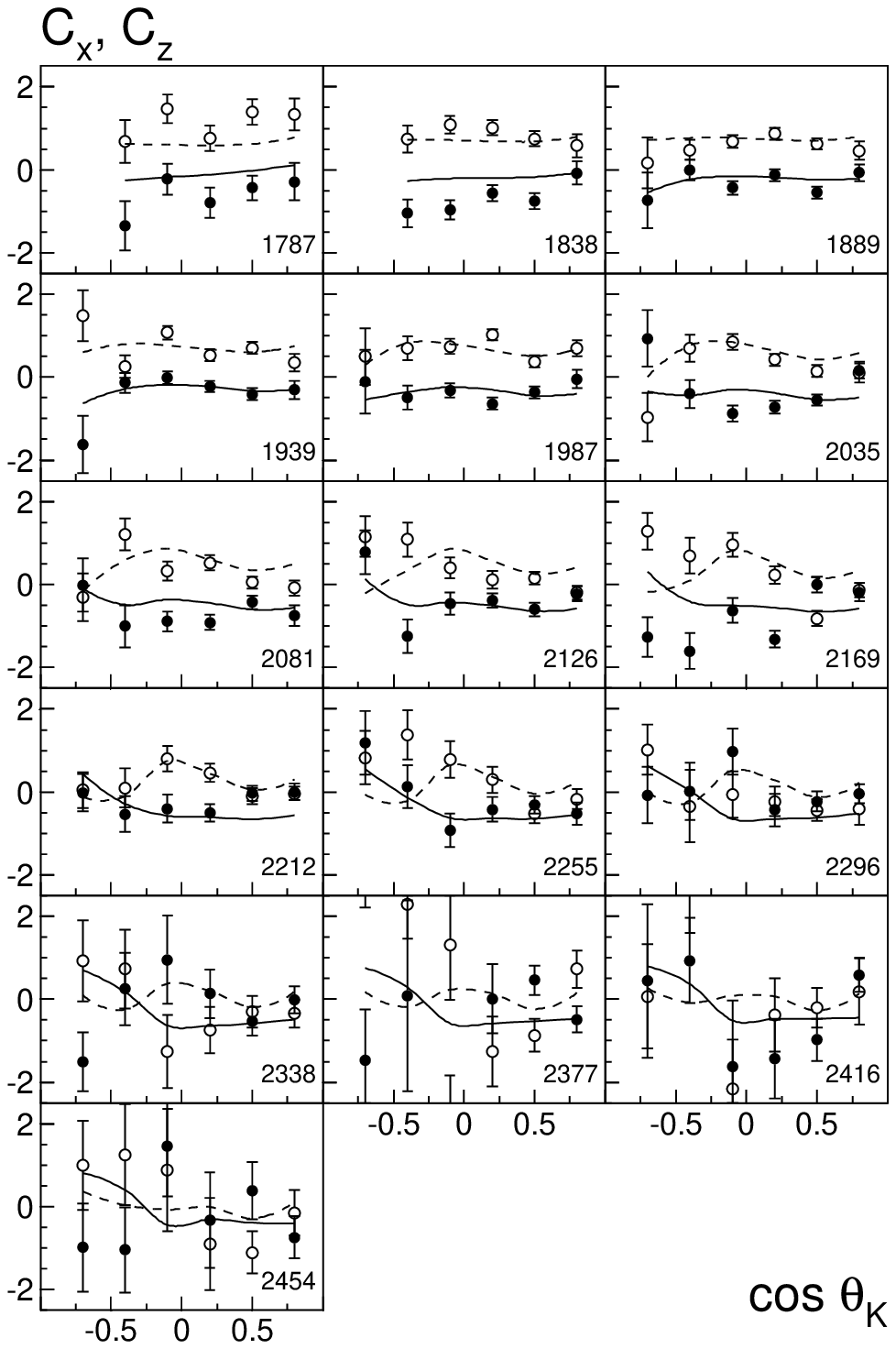,width=0.5\textwidth,height=0.47\textheight}&
\hspace{-4mm}\epsfig{file=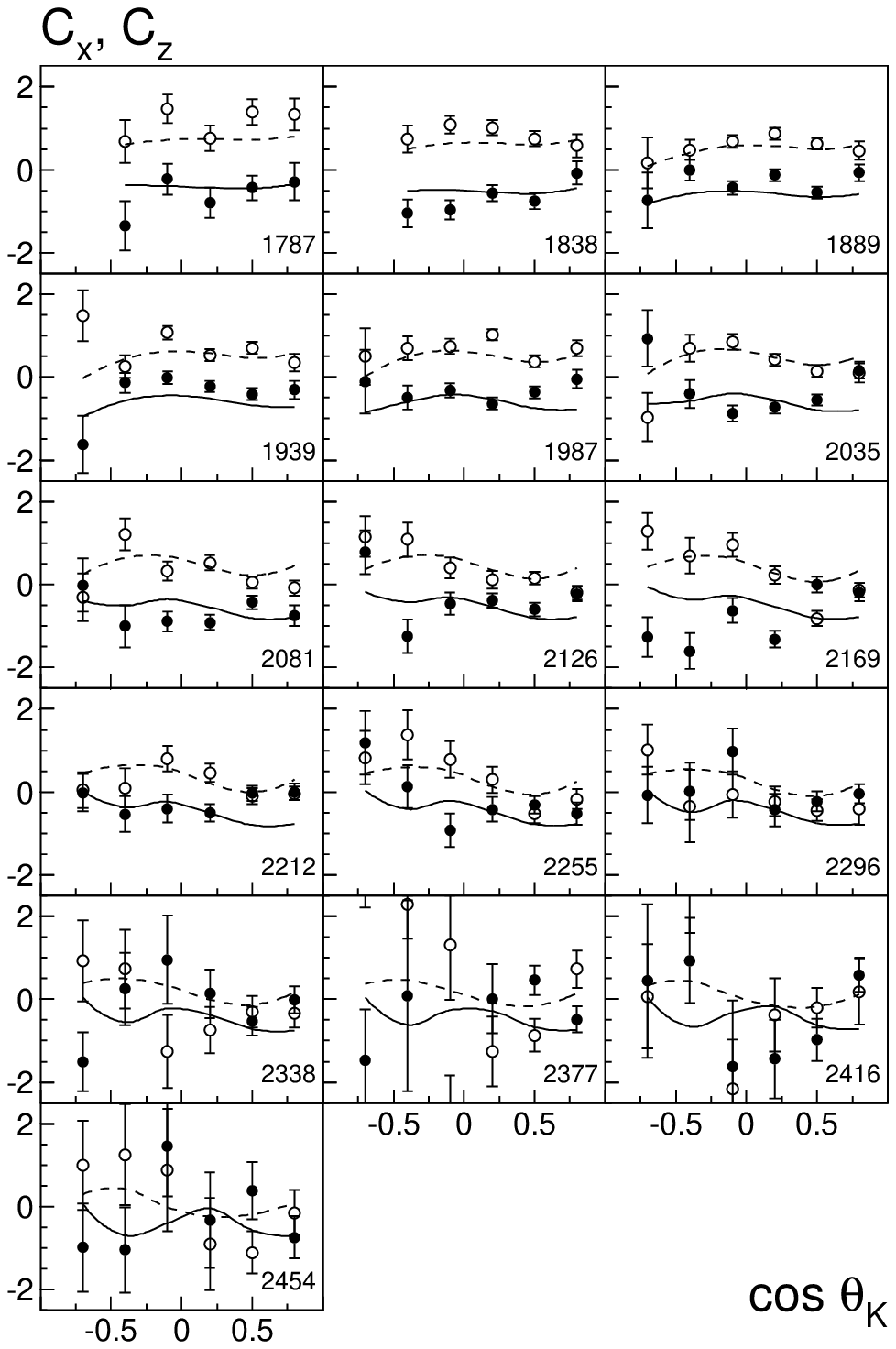,width=0.5\textwidth,height=0.47\textheight}
\end{tabular}
\end{center} \caption{\label{fig:cxcz_klam} Double
polarization observables $C_x$ (black circle) and $C_z$ (open
circle) for $\rm \gamma p\to\KL$ (top panel) and $\rm \gamma
p\to\KS$ (bottom panel) \cite{Bradford:2006ba}. The solid and dashed
curves are results of our fit obtained with solution 1 (left) and
solution 2 (right) for $C_x$ and $C_z$, respectively.}
\end{figure*}

\begin{figure*}[pt]
\begin{tabular}{cc}
\hspace{-2mm}\epsfig{file=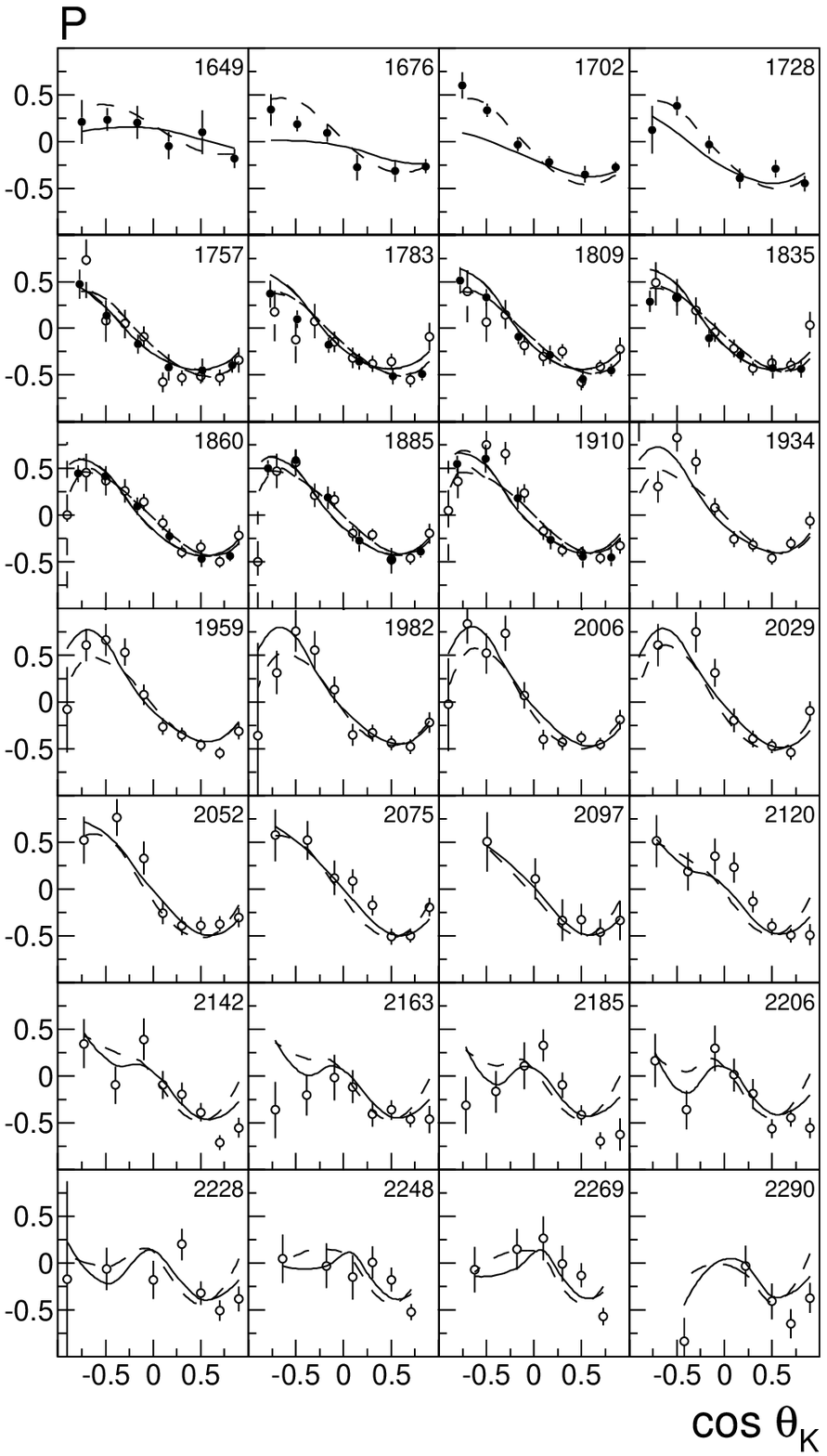,width=0.50\textwidth,height=0.65\textheight}&
\hspace{-4mm}\epsfig{file=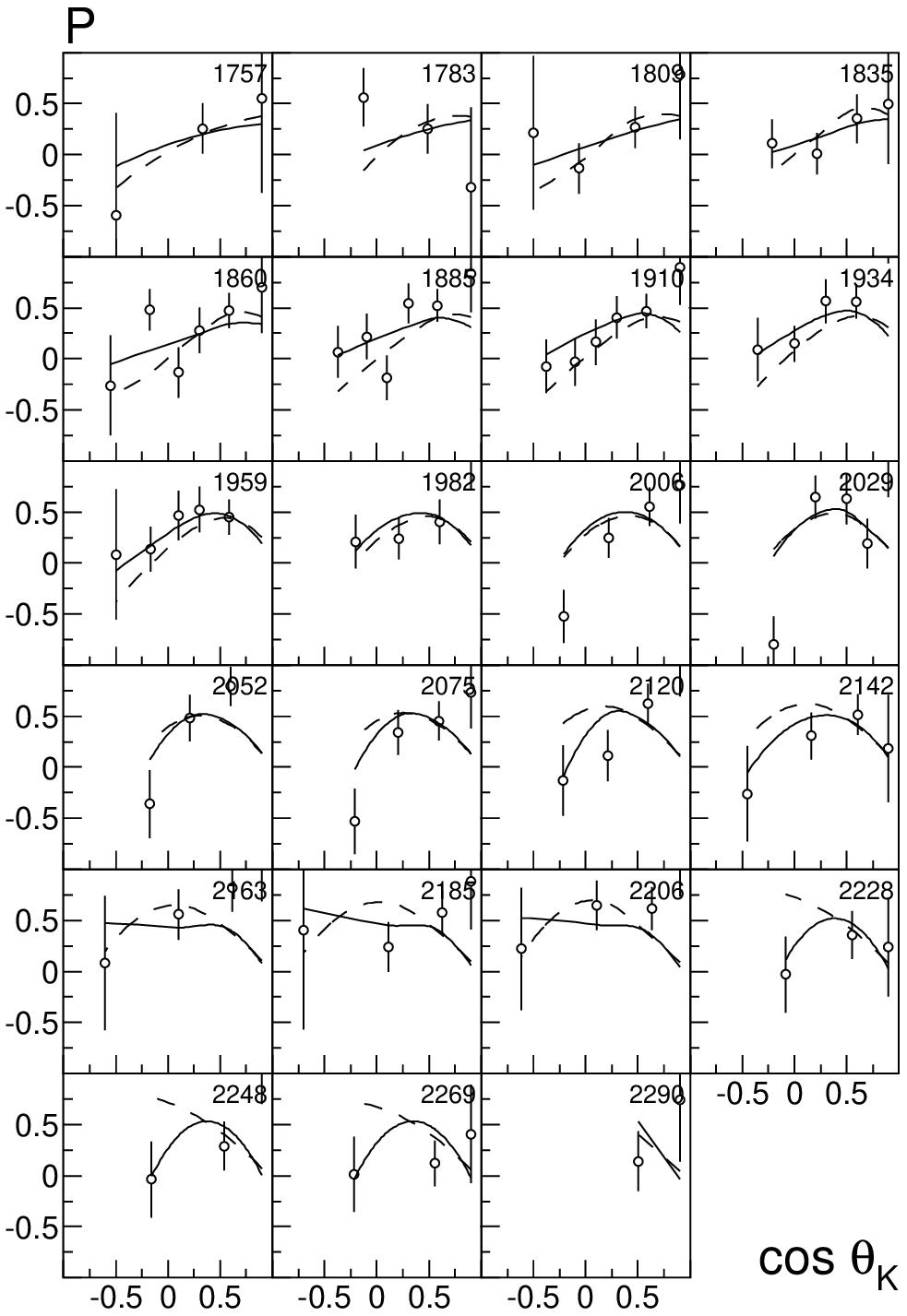,width=0.50\textwidth,height=0.65\textheight}
\end{tabular}
\caption{\label{fig:klam_graal06_p} The recoil polarization
asymmetries as a function of $W$ for $\rm \gamma p\to K^{+}\Lambda$
(left) and $\rm \gamma p\to K^{+}\Sigma^0$ (right) from CLAS
\cite{McNabb:2003nf} (open circle) and GRAAL (black circle)
\cite{Lleres:2007tx}. The solid and dashed curves are the result of
our fit obtained with solution 1 and 2 respectively.} \end{figure*}

 The $\gamma p\to \KL$ differential
cross section was measured recently by CLAS with large statistics
\cite{Bradford:2005pt}. The total cross section in Fig.
\ref{tot_klam} does not show a narrow peak in the $\gamma p
\rightarrow\KL$ cross section at 1700 MeV as suggested by older data
\cite{Glander:2003jw,McNabb:2003nf} but for which we did not find a
physical interpretation in our previous fits
\cite{Anisovich:2005tf,Sarantsev:2005tg}. The SAPHIR data are still
included using a relative normalisation function as described in
\cite{Sarantsev:2005tg}. The total cross section seems to be better
described by solution 1. However, the quality of the description of
the angular distributions is very similar for both solutions (see
Fig. \ref{dcs_klam}); discrepancies in the total cross sections are
due to the extrapolation into regions where no data exist. Hence,
the total cross section cannot be used to favor solution 1 over
solution 2. Note, that the total cross section is calculated as sum
of the measured differential cross sections and the integrated fit
result for the angular region where data are not available.

The total and differential cross sections for $\gamma p\!\to\!\KS$ are
presented in Fig.~\ref{ksig_tot} and in the right panel of
Fig.~\ref{dcs_klam}. The quality of the description is as good as that
obtained without double polarization data taken into account.

The GRAAL collaboration \cite{Lleres:2007tx} measured the $\KL$ and
$\KS$ beam asymmetries in the region from threshold to $W=1906$ MeV.
These data are an important addition to the LEPS data of the beam
asymmetry \cite{Zegers:2003ux} which cover the energy region from
$W=1950$ MeV to $2300$ MeV. Data and fits are shown in Fig.
\ref{fig:klam_graal06_s}.

Fig.~\ref{fig:cxcz_klam} shows the data on $C_x$ and $C_z$ and the
fit obtained with solution 1 and 2. This is the data which gave the
surprising large value for the spin transfer probability from the
circularly polarized photon in the initial state to the final state
hyperon. For both observables a very satisfactory agreement between
data and fit is achieved. Small deviations show up in two mass
slices in the 2.1 GeV mass region. These should however not be
over-interpreted. $C_x^2+C_z^2+P^2$ is constrained by unity; in the
corresponding mass- and $\cos\Theta_K$ bins, $C_z^2$ and the recoil
polarization are sizable pointing at a statistical fluctuation
beyond the physical limits. Of course, a fit must not follow data
into not allowed regions.

Finally, the GRAAL collaboration measured also the
recoil polarization \cite{Lleres:2007tx} for which data from CLAS
\cite{McNabb:2003nf} had been measured in the region from threshold up
to 2300 MeV. These data are reproduced in Fig. \ref{fig:klam_graal06_p}.

\begin{figure}[pt]
\centerline{\epsfig{file=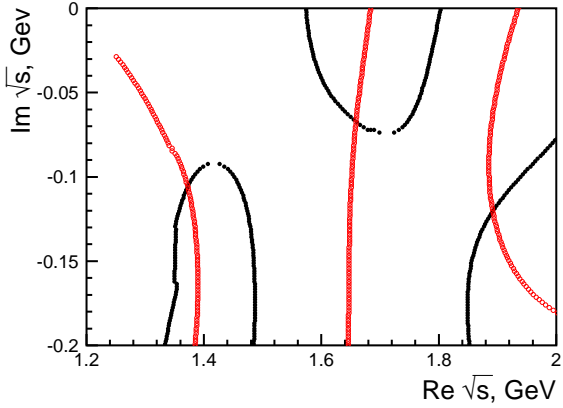,width=0.42\textwidth}}
\caption{\label{contour}Contour lines defined by setting real or
imaginary part of the $P_{11}$ amplitude to zero. The crossing
points define the pole positions. }
\end{figure}

The fits return parameters for each partial waves. From one of the
Tables \ref{Table:p11_km}-\ref{Table:s11_km_1} (and similar tables
for other resonances), the complex amplitude (\ref{amplitude}) can
be calculated using the equations (\ref{psr}) and (\ref{psi}) to
calculated the phase space integral. These complex amplitudes have
poles in the complex $\sqrt s$ plane. Setting the real or imaginary
part of the amplitude to zero defines lines in the $\sqrt s$ plane
which cross at the positions of a pole. This procedure is done for
the different Rieman sheets; the pole closest to the real axis
defines mass and width of the resonance. This plane and the contour
lines defining the pole positions are shown in Fig. \ref{contour}
for the $P_{11}$ partial wave.

The coupling constants $g_a^{(\alpha)}$ defined in (\ref{Pvect})
give the strength of the coupling of a K-matrix pole $\alpha$ to
channel $a$. We are, of course, interested in the couplings of the
resonance $\alpha$ which is defined by its poles of the full
amplitude or T-matrix poles. These couplings, called $\tilde
g_a^{(\alpha)}$ here, are calculated as residues of the T-matrix
amplitudes at the pole positions. For narrow resonances, the
residues are real and can be compared to PDG values. For wide
resonances, the partial widths are functions of $s$, and the
residues acquire phases. To get partial decay widths which can be
compared to other results, we determined the non-relativistic
Breit-Wigner amplitude which reproduced the exact pole position of
the full amplitude; the branching ratios given in the
Table~\ref{Table:p13_all} were calculated at the position of the
Breit-Wigner mass. When this procedure was applied to an amplitude
initially parameterized as Breit-Wigner amplitude, then the original
parameters were reproduced within 1-2 MeV even in the case of very
large couplings and rapidly increasing phase volumes. The method
works very reliably, too, for the K-matrix parameterization, even
with interfering poles. However, the approach is less stable when a
pole is situated close to a threshold and when the pole structure
becomes complicated. In this case, we calculated the residues for
both poles and increased the errors to take into account both
results. In the first solution, the strong interference between the
two lowest $P_{13}$ poles required such an increased systematic
error.

The partial widths of a resonance are related to the coupling $g_a$
by
\be
m_0\Gamma_a=g^2_a \rho(m_0)
\ee
where $\rho(m_0)$ is the phase space defined in (\ref{psr}),
calculated at the real part of the pole position. To minimize the
errors, we determine the decay mode fractions which do not contain
the error of the total width.

\section{Evidence for the \boldmath$N(1900)P_{13}$\unboldmath}

The fits described in this paper used a number of new reactions, and
a variety of different results were obtained which are submitted in
parallel as letter communications. These publications report results
from the same fits to all reactions listed in Table \ref{list_chi}.
The reaction $\gamma p\to p\pi^0\pi^0$ was studied by the CBELSA
collaboration \cite{Thoma:2007}. The analysis returned decays of
baryon resonances in the third resonance region into
$p\pi^0\pi^0$ via different isobars like $\Delta(1232)\pi$,\\
$N(\pi\pi)_{\rm S-wave}$, $N(1440)\pi$, and others. In connection
with precise low-energy data on $\gamma p\to p\pi^0\pi^0$ of the
A2/TAPS collaboration at MAMI, properties of the Roper resonance
were derived \cite{Sarantsev:2007}. The reaction $\gamma p\to
p\pi^0\eta$ \cite{Horn:2007} required introduction of a
$\Delta(1940)D_{33}$ which is suggested to form, jointly with
$\Delta(1900)S_{31}$ and $\Delta(1930)D_{35}$, a triplet of
resonances at a rather low mass, incompatible with quark model
calculations. The new CLAS data on $C_x$ and $C_z$ require
introduction of a $N(1900)P_{13}$ resonance; without it no good
description of the data was reached \cite{Nikonov:2007}. Since this
paper is mainly concerned with hyperon photoproduction, we report
here the main reasons why this resonance is required by the data.

The effect of removing the $N(1900)P_{13}$ resonance from the fit
can be seen in figure 1 of \cite{Nikonov:2007}. The fit quality
changes by $\chi^2_{\rm tot}=1540$ units. The $N(1900)P_{13}$ was
replaced by resonances with other quantum numbers. Replacing it by
an $S_{11}$ \{or $D_{15}$\} state, $\chi^2_{\rm 2b}$ changed by 950
\begin{figure}[pt]
\centerline{\epsfig{file=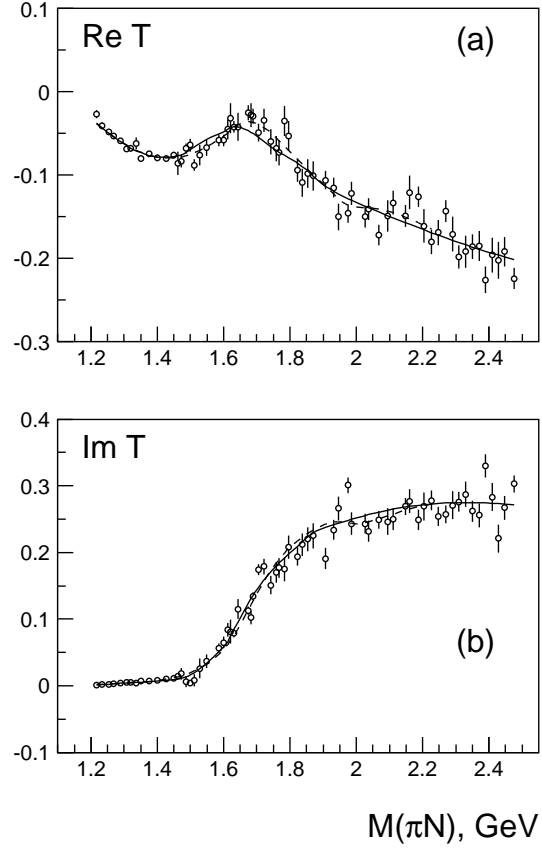,width=0.40\textwidth}}
\caption{\label{pwa_p13} Real (a) and imaginary (b) part of the
$P_{13}$ elastic scattering amplitude \cite{Arndt:2006bf} in
comparison to the fit, solution 1 (solid curve) and solution 2
(dashed curve).}
\end{figure}
\{970\} only. Using $P_{11}$ quantum numbers (instead of $P_{13}$)
gave $\Delta\chi^2_{\rm 2b}=205$ only. An $F_{15}$ state improved
$\chi^2_{\rm 2b}$ marginally; introducing $F_{17}$ and $G_{17}$ did
not improve the fit. A resonance with $P_{33}$ quantum numbers
provided a change in $\chi^2_{\rm 2b}$ which was smaller by a factor
2 than the one found for a $P_{13}$ state.

In a final step, the $P_{13}$ was parameterized as 3-pole 8-channel
K-matrix with $\pi N$, $\eta N$, $\Delta(1232)\pi$ ($P$ and
$F$-waves), $N\sigma$, $D_{13}(1520)\pi$ ($S$-wave), $K\Lambda$ and
$K\Sigma$ channels. This resulted in the fit solutions 1 and 2 which
both are compatible with a large body of data. Both solutions are
compatible with elastic $\pi N$ scattering. Real and imaginary part
of the $P_{13}$ partial wave \cite{Arndt:2006bf} are satisfactorily
described for invariant masses up to 2.4 GeV, see Fig.
\ref{pwa_p13}. From the fit, properties of resonances in the
$P_{13}$-wave were derived. The lowest-mass pole is identified with
the established $N(1720)P_{13}$, the second pole with the badly
known $N(1900)P_{13}$. A third pole is introduced at about 2200 MeV.
It improves the quality of the fit in the high-mass region but its
quantum numbers cannot be deduced safely from the present data base.

The parameters of the two lowest $P_{13}$ poles are collected in
Table~\ref{Table:p13_all}.  The first $P_{13}$ state was found to be
much broader than suggested by most other analyses
\cite{Yao:2006px}. However, the only analysis taking $N\pi\pi$ data
into account gives a width of $(380\pm 180)$\,MeV
\cite{Manley:1992yb}. The most recent analysis of elastic scattering
data \cite{Arndt:2006bf} gave a 355\,MeV width. The elastic width of
the $N(1720)P_{13}$ ($\approx 45$\,MeV) is even narrower than the
$N(1680)F_{15}$ elastic width $\approx 85$\,MeV). Given the large
spread of pole positions reported in \cite{Yao:2006px}, we do not
think that our result is in conflict with previous work.

In the first solution, the double structure in the $P_{13}$ partial
wave (see Fig. \ref{tot_klam}a) is due to a strong interference
between the first and the second pole. If the structure is fitted
with one pole, the pole must have a rather narrow width. The
$N(1720)P_{13}$ couples strongly to $\Delta(1232)\pi$ and, in the
second solution, also to the $D_{13}(1520)\pi$ channel. The
$D_{13}(1520)\pi$ threshold is close to its mass and creates a
double pole structure which makes difficult the definition of
helicity amplitudes and of decay partial widths. The method used
here is described in \cite{Nikonov:2007}.

\begin{table}[pt]
\caption{\label{Table:p13_all} Properties of the two lowest $P_{13}$
resonances for both solutions. The masses, widths are given in MeV,
the branching ratios in \% and helicity couplings in 10$^{-3}$
GeV$^{-1/2}$. The helicity couplings and phases were calculated as
residues in the pole position. \vspace{2mm}}
\begin{center}
\renewcommand{\arraystretch}{1.4}
\begin{tabular}{lcccc}
\hline\hline
&\multicolumn{2}{c}{Solution 1}&\multicolumn{2}{c}{Solution 2}\\
\hline
$M_{pole}$               &\hspace{-4mm}$1640\pm 80 $       &\hspace{-3mm}$1870\pm 15$       &\hspace{-3mm}$1630\pm 60 $       &\hspace{-3mm}$1960\pm 15$\\
$\Gamma_{tot}^{pole}$    &\hspace{-4mm}$ 480\pm 80 $       &\hspace{-3mm}$ 170\pm 30$       &\hspace{-3mm}$ 440\pm 60 $       &\hspace{-3mm}$ 195\pm 25$\\
\hline
$A_{1/2}$                &\hspace{-4mm}$ 140\pm 80$        &\hspace{-3mm}$ -(10\pm 15)$       &\hspace{-3mm}$160\pm 40$         &\hspace{-3mm}$-(18\pm 8) $\\
$\varphi_{1/2}$          &\hspace{-4mm}$-(10\pm 15)^\circ$ &\hspace{-3mm}  --               &\hspace{-3mm}$ (10\pm 15)^\circ$ &\hspace{-3mm}$ -(40\pm 15)^\circ $\\
$A_{3/2}$                &\hspace{-4mm}$ 150\pm 80$        &\hspace{-3mm}$-(40\pm 15)$      &\hspace{-3mm}$ 70\pm 30$         &\hspace{-3mm}$-(35\pm 12)$\\
$\varphi_{3/2}$          &\hspace{-4mm}$-(40\pm 30)^\circ$ &\hspace{-3mm}$ (30\pm 25)^\circ$&\hspace{-3mm}$  (0\pm 20)^\circ$ &\hspace{-3mm}$-(40\pm 15)^\circ$\\
\hline
${\rm Br}_{N\pi}$        &\hspace{-4mm}$ 8\pm 4$           &\hspace{-3mm}$ 5\pm  3$         &\hspace{-3mm}$18\pm 5$           &\hspace{-3mm}$ 6\pm 3$\\
${\rm Br}_{N\eta}$       &\hspace{-4mm}$14\pm 4$           &\hspace{-3mm}$20\pm  8$         &\hspace{-3mm}$10\pm 2$           &\hspace{-3mm}$15\pm 3$\\
${\rm Br}_{K\Lambda}$    &\hspace{-4mm}$16\pm 6$           &\hspace{-3mm}$15\pm  5$         &\hspace{-3mm}$ 7\pm 2$           &\hspace{-3mm}$12\pm 3$\\
${\rm Br}_{K\Sigma}$     &\hspace{-4mm}$ <2    $           &\hspace{-3mm}$22\pm  8$         &\hspace{-3mm}$<1     $           &\hspace{-3mm}$ 8\pm 2$\\
${\rm Br}_{\Delta\pi(P)}$&\hspace{-4mm}$54\pm 10$          &         &\hspace{-3mm}$36\pm 6$           &\\
${\rm Br}_{\Delta\pi(F)}$&\hspace{-4mm}$ 2\pm 2$           &         &\hspace{-3mm}$18\pm 5$           &\\
${\rm Br}_{D_{13}\pi}$   &\hspace{-4mm}$2\pm 2$            &         &\hspace{-3mm}$ 5\pm 3$           &\\
${\rm Br}_{N\sigma}$     &\hspace{-4mm}$ 4\pm 2$           &         &\hspace{-3mm}$ 4\pm 2$           &\\
${\rm Br}_{Add}$         &\hspace{-4mm}$<2$ &\hspace{-3mm}$38\pm 12$         &\hspace{-3mm}$ 2\pm 2$&\hspace{-3mm}$60\pm 6$\\
\hline\hline
\end{tabular}
\renewcommand{\arraystretch}{1.0}
\end{center}
\end{table}

The $N(1900)P_{13}$ is of special interest for baryon spectroscopy.
It belongs to the two-star positive-parity $N^*$ resonances in the
1900 - 2000 MeV mass interval -- $N(1900)P_{13}$, $N(2000)F_{15}$,
$N(1990)F_{17}$ -- which cannot be assigned to quark-diquark
oscillations \cite{Santopinto:2004hw} when the diquark is treated as
point-like object with zero spin and isospin. At the present status
of our knowledge on baryon excitations, most four-star and
three-star baryon resonances can be interpreted in a simplified
model describing baryons as being made up from a diquark and a
quark. Evidence for the $N(1900)P_{13}$ has been communicated
already in a letter publication \cite{Nikonov:2007}. The
$N(2000)F_{15}$ is included in the analysis as well; it is a further
two-star  $N^*$ resonances which cannot be assigned to quark-diquark
oscillations. The evidence for this state from this analysis is,
however, weaker. The $N(1840)P_{11}$ state (which we now find at
1880\,MeV) could be the missing partner of a super-multiplet of
nucleon resonances having -- as leading configuration -- intrinsic
orbital angular momentum $\ell=2$ and a total quark spin $s=3/2$.
These angular momenta couple to a series $J=\frac 12,\frac 32,\frac
52,\frac 72$. Yet in this analysis, there was no need to introduce
$N(1990)F_{17}$.
\section{Discussion}

Triggered by the measurement of the spin transfer coefficients $C_x$
and $C_z$ we have refitted data on single $\pi$, $\eta$, $K^0$ and
$K^+$ photoproduction. Besides $C_x$ and $C_z$ the new data on
unpolarised differential cross section for $\gamma N\to \KL$,
$\gamma N\to \KS$, and $K^0 \Sigma^+$ photoproduction, and double
pion production data were added to the combined analysis. The refit
was motivated by the bad prediction of the spin transfer
coefficients with our previous partial wave analysis. All data sets
can be described well after introducing a $N(1900)P_{13}$ resonance.
Its mass and width are estimated to $1915\pm 50$ MeV and $180\pm 50$
MeV, respectively. This result covers the two K-matrix solutions
found here: in the first one, the pole position of the second
$P_{13}$ state is located at $1870-i\,85$ MeV and in the second
solution at $1960-i88$ MeV. The reason for the ambiguity is likely
connected with the existence of a $P_{11}$ state with similar mass
and width.

Even though the description of all distributions is very reasonable,
the two solutions have remarkably different isobar contributions. In
solution 1, the $P_{13}$ partial wave shows a significant double
structure (not present in solution 2).  The $S_{11}$ wave is much
stronger in solution 2.

The new $P_{13}$ state \cite{Nikonov:2007} also improves the
description of the $\gamma p\to \KS$ reaction. However the effect
from introducing this state is much smaller here. Actually, in our
previous analysis, the double polarization data of this channel were
already described much better than those for $\gamma p\to K\Lambda$
(see figures in \cite{Bradford:2006ba}); a fully satisfactory
description was already achieved after a slight readjustment of the
fit parameters. Nevertheless, the $P_{13}$ state definitely improved
the description and provided a noticeable signal in the $\gamma p\to
\KS$ total cross section (see Fig. \ref{ksig_tot}). Differential and
total cross sections had already been described successfully when a
$P_{11}$ state was introduced at 1840 MeV \cite{Sarantsev:2005tg}.
When both states, $P_{11}$ and $P_{13}$, were introduced, they share
about equal contributions to this cross section. The statistical
significance for two states was however not convincing. So, at the
end, only one resonance was introduced in \cite{Sarantsev:2005tg};
the likelihood favoured $P_{11}$ quantum numbers.

The main contribution to $\gamma p\to K^+\Sigma^0$ now originates
from $K$-exchange while we had found a larger $K^*$-exchange
contribution in  \cite{Sarantsev:2005tg}. The preference for Kaon
exchange gives a natural explanation for the small $\gamma p\to
K^0\Sigma^+$ cross section where Kaon exchange is forbidden. The
$P_{13}$ partial wave provides a moderate contribution but helps to
describe data in the 1870 MeV mass region.

Although a qualitatively good description of all fitted observables
was obtained, both solutions have some local problems. Due to the
larger statistics, problematic deviations are more easily seen in
the $\KL$ distributions.

The first solution does not describe the $\KL$ recoil polarization
at backward angles at the energy around 1700 MeV. The second
solution describes this region better. The second solution, in turn,
has some problem in the $C_x$ and recoil polarization at higher
energies, in the 2100 MeV region. The description can be improved in
two ways; for the first solution, introduction of an additional
state in the 1800 MeV region solves the problem. This could be
either a $S_{11}$, a $D_{15}$ or $P_{33}$ resonance. Yet, we are not
sure that these additional states can be identified well with the
present quality of the data.  Hence we postpone the identification
of the weaker signals until new data are available. The main result
of the present analysis is that a qualitatively and quantitatively
satisfactory description of the fitted data can be obtained by
introduction of a new $P_{13}$ relatively narrow state in the region
1885 MeV (solution 1) or in the region 1970 MeV (solution 2).

The partial wave analysis presented here demonstrates that the CLAS
findings, that the $\Lambda$ (and $\Sigma$) hyperons are produced
100\% (or 80\%) polarized, can be described quantitatively in the
conventional picture where intermediate resonances strongly contribute to the
dynamics of the reactions. Even in the case of large non-resonant contribution
baryon resonances still play an important role in
the dynamics of the process.
On the other hand, the analysis also shows
that even data sets comprising various high-statistics differential cross
sections, beam, target and recoil asymmetries, double polarization
observables,  and data which resolve the two isospin contributions
(by a simultaneous analysis of the $p\pi^0$ and $n\pi^+$,
the $K^+\Sigma^0$ and the $K^0\Sigma^+$ as well as the isospin
selective $p\eta$ and $K^+\Lambda$ channels) are still
not yet sufficient to converge into a unique solution. Systematic
measurements with further double polarization observables -- as being
planned and carried out at several laboratories -- are urgently needed.

\section*{Acknowledgements}

The work was supported by the DFG within the SFB/TR16 and by a FFE
grant of the Research Center J\"ulich. U. Thoma thanks for an Emmy
Noether grant from the DFG. A.~Sa\-ran\-tsev gratefully acknowledges
the support from Russian Science Support Foundation. This work is
also supported by Russian Foundation for Basic Research
07-02-01196-a and Russian State Grant Scientific School
\\ 5788.2006.2.

\end{document}